\documentclass{aa}  
\usepackage{graphicx}
\usepackage{txfonts}
\usepackage{textcomp}

\usepackage{natbib}
\bibpunct{(}{)}{;}{a}{}{,} % to follow the A&A style

\catcode`\@=11
\def\gsim{\ifmmode{\mathrel{\mathpalette\@versim>}}
   \else{$\mathrel{\mathpalette\@versim>}$}\fi}
\def\lsim{\ifmmode{\mathrel{\mathpalette\@versim<}}
    \else{$\mathrel{\mathpalette\@versim<}$}\fi}
\def\@versim#1#2{\lower 2.9truept \vbox{\baselineskip 0pt \lineskip
    0.5truept \ialign{$\m@th#1\hfil##\hfil$\crcr#2\crcr\sim\crcr}}}
\catcode`\@=12

\usepackage{xcolor}
\newcommand{\noun}[1]{\textsc{#1}}

\begin{document} 
\title{Homogeneous metallicities and radial velocities for Galactic globular clusters
\thanks{Based on observations taken with ESO telescopes at the 
   La Silla Paranal Observatory under programme ID 089.D-0493(B).}
\thanks{Table 3 is available in electronic form
at the CDS via anonymous ftp to cdsarc.u-strasbg.fr (130.79.128.5)
or via http://cdsweb.u-strasbg.fr/cgi-bin/qcat?J/A+A/}
}
\subtitle{II. New CaT metallicities for 28 distant and reddened globular clusters}

\author{
 S. V\'{a}squez\inst{1} 
 \and
 I. Saviane\inst{2}
 \and
 E. V. Held\inst{3}
 \and 
 G. S. Da Costa\inst{4}  
 \and
 B. Dias\inst{2}
 \and 
 M. Gullieuszik\inst{3} 
 \and 
 B. Barbuy\inst{5} 
 \and 
 S. Ortolani\inst{6,3}
 \and
 M. Zoccali\inst{7,8}
}

\institute{
Museo Interactivo Mirador, Direcci\'{o}n de Educaci\'{o}n, Av. Punta Arenas 6711, La Granja, Santiago, Chile
\and
European Southern Observatory, Alonso de Cordova 3107, Santiago, Chile
\and
INAF -- Osservatorio Astronomico di Padova, vicolo Osservatorio 5, 35122 Padova, Italy
\and 
Research School of Astronomy \& Astrophysics, Australian National University, Mt Stromlo Observatory, via Cotter Rd, Weston, ACT 2611, Australia
\and 
{Universidade de S\~ao Paulo, Rua do Mat\~ao 1226, S\~ao} Paulo, 05508-900, Brazil
\and 
Dipartimento di Fisica e Astronomia, Universit\`a degli Studi di Padova, vicolo dell'Osservatorio 3, 35122 Padova, Italy
\and
Instituto de Astrof\'{i}sica, Facultad de F\'{i}sica, Pontificia Universidad Cat\'{o}lica de Chile, 
Av. Vicu\~na Mackenna 4860, Santiago, Chile.
\and
Millennium Institute of Astrophysics, 
Av. Vicu\~na Mackenna 4860, Santiago, Chile.
\\
%\email{svasquez@astro.puc.cl} 
}

\date{Received; accepted}

% \abstract{}{}{}{}{} 
% 5 {} token are mandatory
 
\abstract {
Although the globular clusters in the Milky Way have been  studied for a long time, a significant fraction of them lack  homogeneous metallicity and radial velocity measurements. 
In an earlier paper we presented the first part of a project to obtain metallicities and radial velocities of Galactic globular clusters from multiobject spectroscopy  of their member stars using the ESO Very Large Telescope. 
In this paper we add metallicities and radial velocities for a new sample of 28 globular clusters, including in particular globular clusters in the MW halo and the Galactic bulge. Together with our previous results, this study brings the number of globular clusters with homogeneous measurements to $\sim {69}$\% of  those listed in the W. Harris' catalogue. As in our previous work, 
we have used the \ion{Ca}{ii} triplet lines to derive metallicities and radial velocities.  For most of the clusters in this study, this is the first analysis based on spectroscopy of individual member stars.  
The metallicities derived from the \ion{Ca}{ii}  triplet are then compared to the results of our parallel study based on spectral fitting in the optical region and the implications for different calibrations of the \ion{Ca}{ii} triplet  line strengths are discussed. 
{We also comment} on some interesting clusters and investigate the presence of  an abundance spread in the globular clusters here. A hint of a possible intrinsic spread is found for NGC\,6256, which therefore appears to be a good candidate {for further study}. 
%
% percentage considering 157 clsuters from H10:
%%% R97a analysed 52 clusters; R97b compiled W' from literature and averaged them for 71 clusters (45%); S12 analysed 20 clusters, but 4 are in common with R97b (71+16=87 = 55%); S16 analysed 28 clusters, but 6 are in common with R97b and none in common with S12 (87+22=109 = 69%.)
% I didn't check the 60% of clusters analysed for the first time (17 of 28...)
}


\titlerunning{Metallicities and Radial Velocities for Galactic Globular Clusters - II}

   \maketitle
%
%________________________________________________________________
\section{Introduction}

Globular clusters (GCs) were considered as simple stellar populations for a long time, with most of their parameters derived from heterogeneous methods and data, largely driven by the technology available at the time when the clusters were studied. 
{Recently,} with the acquisition of extremely precise photometry from Hubble Space Telescope (HST) imagers and spectroscopy from large telescopes, the idea of simplicity has been  replaced by a more complex view of their star forming history \citep[see, e.g.,][]{gratton+2012}.
{Closely linked to that history is the dynamical evolution of clusters, which is another area of study that benefits from homogeneous samples and reduction techniques. Data sets like these can be used as input to N-body simulations, and the resulting dynamical parameters allow comparative studies such as that recently published in \citet{baumgardt+2018}.}

Improvements in our understanding of GC evolution came also from the introduction of large spectroscopic studies, which enabled measuring homogeneous abundances for large samples of GC stars. 
The largest sample of metallicities measured with the same method comes from the study of equivalent widths of the \ion{Ca}{II}  triplet (CaT) lines in the spectra of GC stars by \citet{rutledge+1997} who  determined metallicities for 52 clusters and made a literature compilation on their scale for a total of 71 clusters.
The second largest sample  is represented by integrated-light spectroscopic studies such as those of  \citet[][ZW84]{zinn+1984} and \citet{armandroff+1988} which account for an additional $\sim20\%$ of the metallicity measurements for GCs. Another approach to metallicity homogenisation was followed by \citet{carretta+2009} using their high-resolution spectroscopic analysis of 19 GCs to set a new metallicity scale based on \ion{Fe}{i} lines.  Transformations to the new scale were given for 114 GCs from \citet{harris+1996}.

The lack of homogeneous measurements currently mostly affects the outer-halo (distant) and bulge (highly extincted) globular clusters, which give important constraints on the Milky Way formation. 
This motivated our group 
{to increase the sample of GCs with metallicities determined in a consistent way.  The project is based on medium-resolution spectroscopy of individual red giant branch (RGB) stars in a sample of clusters which complements that of \citet{rutledge+1997}. 
The spectra were obtained with FORS2 at the ESO VLT observatory in two spectral intervals, the CaT region and the green region that is rich in metal absorption lines.}

In Paper~I \citep[][S12]{saviane+2012} we presented the methods and analysis of spectra in the CaT region for a first set of clusters from the original sample, including eight {calibration} GCs and 20 programme clusters, four of which are in common with \citet{rutledge+1997}.  {The use of spectra at the CaT} is particularly useful for targets near the Galactic plane or towards the Milky Way bulge because of  the lower extinction in the far-red spectral range.
A new CaT metallicity calibration based on the \citet[][hereafter C09]{carretta+2009} abundance scale was derived and used to measure the metallicity of the programme clusters. In addition, the clusters from  \citet{rutledge+1997} were included, converted to the new scale. 

In a parallel project, new metallicity measurements were obtained for a largely overlapping GC sample from {full spectral fitting} of medium resolution spectra in the visible region \citep{dias+2015,dias+2016}. The new metallicity scale, independent of previous empirical calibrations, appears to be consistent both with abundances from high-resolution spectroscopy and photometric metallicities \citep{cohen+17}. 
On the basis of the new scale, and taking advantage of an increased sample of metal-rich clusters,  \citet{dias+2016eso} {generated a homogeneous compilation} of metallicities for 152 GCs (97\% of the total) from different sources. 

This paper increases our homogeneous database by adding new CaT measurements for a further set of 28 GCs, which in combination with those published in \citet{saviane+2012}  provide metallicity estimates for ${\sim 69\%}$ 
of the Milky Way globular clusters in the \citet{harris+1996}\footnote{Hereafter all references to the catalogue refer to the 2010 on-line version \citep[][]{harris+2010}, unless otherwise specified.} catalogue. 
The [Fe/H] estimates from the CaT are then compared to the GC metallicities independently obtained from the green spectral interval \citep{dias+2015,dias+2016}. 

This paper is organised as follows: in Sect.~2  the observations and data reduction are presented; in Sect.~3, radial velocities  are measured and compared with literature values; in Sect.~4 several CaT EW calibration relations are discussed and metallicities are estimated.  Section~5  provides  comments for some interesting clusters. Finally, our results are summarised in Sect.~6. 

\begin{figure}[t]
\centering
% \includegraphics[angle=0,width=1.0\columnwidth]{plots/fig1.eps}
%> \includegraphics[angle=0,width=1.0\columnwidth]{newplots/fig1_ivo.ps}
 \includegraphics[angle=0,width=1.0\columnwidth]{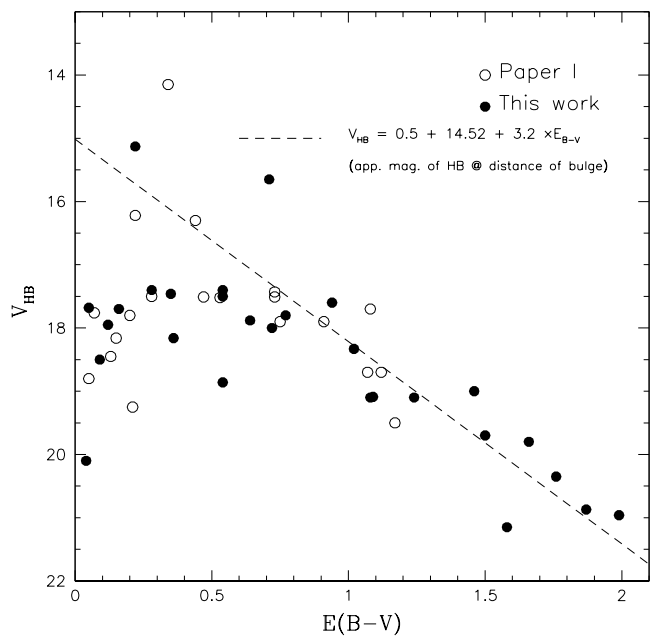}
% {\includegraphics[angle=0,width=1.0\columnwidth]{plots_new/01_reddening_VBH.eps}}
 \caption{Reddening and visual apparent magnitude at the level of the
   horizontal branch for the globular clusters observed in
   \citet{saviane+2012} and this work. Data are from the Harris catalogue.
   The dashed line shows the $V$ magnitude of the horizontal branch 
     (assuming $V_{0,\rm{HB}}=0.5$) for a cluster at the distance of the
     bulge (8~kpc) and subject to different degrees of extinction.}
  \label{reddening_plot}
  \end{figure}

%__________________________________________________________________
\section{Observations and Data Reduction}

\begin{table}
\caption{Observing log including exposure time, reddening and metallicity source for the 28 programme  clusters.  \label{tab:Observations-log}}
\begin{centering}
\begin{tabular}{llll}
% &  &  &  \tabularnewline
\hline 
\hline 
Cluster   & $t_{{\rm exp}}$ (s)  & $E(B-V)$  & {[}Fe/H{]}\tabularnewline
\hline 
%%% &  &  & \tabularnewline
\noalign{\smallskip}
BH 176   & $1\times240$ & 0.54 &    \tiny{(Phot)/  PS03}                 \\  % PS03}                 \\
Djorg 1  & $2\times1560$& 1.58 &    \tiny{(Phot)/  VFO10}                \\  % VFO10}                \\
Djorg 2  & $1\times60$  & 0.94 &    \tiny{(Phot)/  VFO10}                \\  % VFO10}                \\
NGC 5634 & $1\times60$  & 0.05 &    \tiny{(i)/     ZW84 (Q39)}           \\  % ZW84 (Q39)}           \\
NGC 5694 & $1\times240$ & 0.09 &    \tiny{(i)/     AZ88, LLC06}          \\  % Car09, AZ88, LLC06}   \\
NGC 5946 & $1\times60$  & 0.54 &    \tiny{(i)/     AZ88}                 \\  % Car09, AZ88}          \\
NGC 6256 & $1\times360$ & 1.09 &    \tiny{(i)/     SF04}                 \\  % SF04}                 \\
NGC 6284 & $1\times60$  & 0.28 &    \tiny{(i)/     ZW84 (Q39)}           \\  % ZW84 (Q39)}           \\
NGC 6316 & $1\times60$  & 0.54 &    \tiny{(i)/     AZ88}                 \\  % Car09, AZ88}          \\
NGC 6352 & $1\times60$  & 0.22 &    \tiny{(h)/     F91, FPJ09}           \\  % Car09, F91, FPJ09}    \\
NGC 6355 & $1\times120$ & 0.77 &    \tiny{(i)/     ZW84 (Q39)}           \\  % Car09, ZW84 (Q39)}    \\
NGC 6366 & $1\times60$  & 0.71 &    \tiny{(CaT)/   DA95, DS89}           \\  % Car09, DA95, DS89}    \\
NGC 6401 & $1\times120$ & 0.72 &    \tiny{(Fe)/    M95a}                 \\  % Car09, M95a}          \\
NGC 6426 & $1\times200$ & 0.36 &    \tiny{(i)/     ZW84 (Q39)}           \\  % ZW84 (Q39)}           \\
NGC 6453 & $1\times120$ & 0.64 &    \tiny{(h)/     VFO10}                \\  % Car09, VFO10}         \\
NGC 6517 & $1\times420$ & 1.08 &    \tiny{(Fe)/    M95a}                 \\  % Car09, M95a}          \\
NGC 6539 & $1\times180$ & 1.02 &    \tiny{(i)/     SF04, O\&\&05}        \\  % Car09, SF04, O\&\&05} \\
NGC 6749 & $1\times780$ & 1.50 &    \tiny{(Phot)/  KHM97}                \\  % KHM97}                \\
NGC 6864 & $90/180/60$  & 0.16 &    \tiny{(h)/     Car09}                \\  % Car09}                \\
Pal10   & $1\times660$ & 1.66 &    \tiny{(Phot)/  KH97}                 \\  % KH97}                 \\
Pal11   & $60/180$     & 0.35 &    \tiny{(CaT)/   ADZ92, DA95}          \\  % Car09, ADZ92, DA95}   \\
Pal14   & $1\times1100$& 0.04 &    \tiny{(CaT)/   ADZ92}                \\  % Car09, ADZ92}         \\
Pal6    & $1\times550$ & 1.46 &    \tiny{(h)/     LC02, SF04, LCB04}    \\  % LC02, SF04, LCB04}    \\
Terzan 1 & $1\times1200$& 1.99 &    \tiny{(i)/     AZ88, I\&\&02}        \\  % AZ88, I\&\&02}        \\
Terzan 2 & $1\times1100$& 1.87 &    \tiny{(i)/     AZ88, SF04}           \\  % AZ88, SF04}           \\
Terzan 8 & $3\times120$ & 0.12 &    \tiny{(CaT)/   DA95, MWM08}          \\  % DA95, MWM08}          \\
Terzan 9 & $1100/100$   & 1.76 &    \tiny{(Phot)/  VFO10}                \\  % VFO10}                \\
Ton 2    & $1\times360$ & 1.24 &    \tiny{(Phot)/  BOB96}                \\  % BOB96}                \\
% &  &  & \tabularnewline
\hline 
\end{tabular}
\par\end{centering}
\tablefoot{The metallicity source from Harris catalogue \citep{harris+1996} is coded as (method)/references, where the reference acronym can be found in the catalogue bibliography list. The methods are  (h)igh resolution, (i)ntegrated light, medium-resolution EW of (Fe) lines, (CaT) method or (Phot)ometric studies.} 
\end{table}

The complete cluster programme of this project contains 56 clusters, selected from the catalogue of Galactic globular clusters published by \citet[version February 2003]{harris+1996}. 
{The observations and data reduction followed the procedure described in \citet{saviane+2012},
where we presented our CaT results for the clusters observed in ESO Period~77. }
In order to complete the project, observing time using FORS2 mounted on the UT1-Antu telescope was assigned in ESO P89. Observations were performed in MOS mode for all clusters, using 19 movable slitlets to obtain multislit spectra. The 1028z+29 grism and the OG590+32 order-blocking filter were set, obtaining a spectral coverage 
$\sim 7700$--9500 \AA\  at a scale of 0.85~\AA{}\/ per (binned) pixel. The observing log is given in Table~\ref{tab:Observations-log}. 
As shown in Fig.~\ref{reddening_plot}, the clusters analysed in this work cover the faint and redder part of our full cluster sample. 
{A target list of RGB stars was selected in a narrow range from 1 mag below to 3 mag above the horizontal branch (HB) from colour-magnitude diagrams (CMDs) obtained from the pre-imaging observations in the $V$, $ I$ bands, and used to define the MOS slitlets configuration.} To this aim, we set  a magnitude threshold depending on the cluster and generally fainter than the limit $V=20$ used in \citet{saviane+2012}. 

All the spectra were extracted using the FORS2 pipeline version 4.9.
{The pipeline reduces each target spectrum using daytime calibration frames to compute a wavelength calibration and distortion correction for each slit. The wavelength calibration is improved by aligning the solutions to the position of the sky lines ($skyalign$ option).}
%-
The individual spectra were corrected for bias, flat field and local sky background. The final spectra were extracted using the \citet{horne+1986} optimal extraction and average combined (after radial velocity correction) when more than one spectrum was observed.

%__________________________________________________________________
\section{Radial velocities}

\begin{figure}
\centering
%>  {\includegraphics[angle=0,width=1.0\columnwidth]{plots/02_NGC5946_selection.eps}}
  {\includegraphics[angle=0,width=1.0\columnwidth]{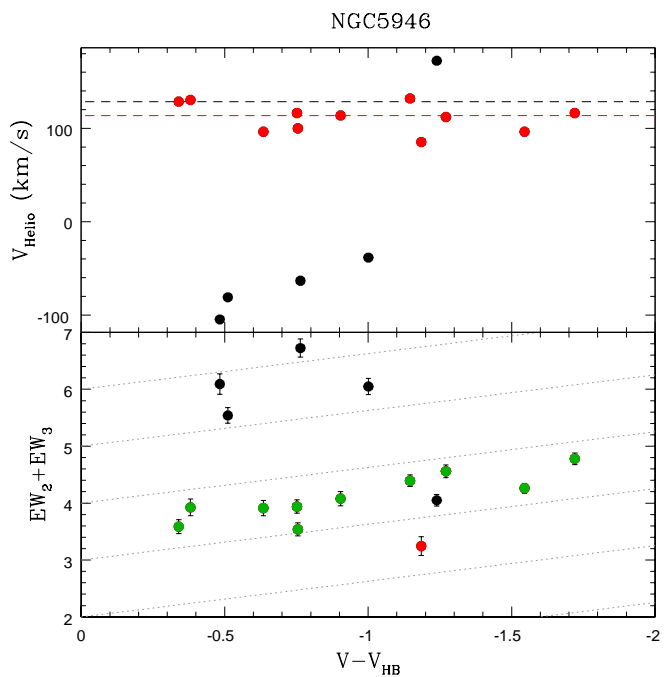}}
  \caption{Cluster member selection for NGC\,5946. The upper panel shows in red the stars selected on the basis of radial velocity criteria, while black dots represent non-members. The mean radial velocity derived from the selected stars is shown as a red dashed line. This value is $\sim15$ km s$^{-1}$ lower than the radial velocity in the \citet{harris+1996} catalogue (black dashed line).  The lower panel shows the selection based on the line strength of the \ion{Ca}{ii} triplet lines. For member selection we used {isoabundance lines with a fixed slope as in \citet{saviane+2012} (dotted lines).} The green dots are stars accepted as cluster members on the basis of both criteria. Only one radial-velocity member  was excluded  from the sample because of its line strength (red dot).}
  \label{selection}
\end{figure}

\begin{table}
\caption{Mean radial velocities of globular clusters  \label{tab:RV}}
\begin{centering}
\begin{tabular}{lrrrrrr}
\hline 
\hline 
Cluster   & $N$  & $v_{\rm r}$ & err  & $v_{\rm r}$ & err & $\Delta v_{\rm r}$\tabularnewline
 &  &  &  & (lit.) &  & \tabularnewline
\hline 
%  &  &  &  &  &  & \tabularnewline
\noalign{\smallskip}
BH 176   & 11 & 0.1    & 0.7 &   0.0   & 15   &  0.1\\ 
Djorg 1  & 3  & -358.1 & 0.7 & -362.4 & 3.6 & 4.2   \\ 
Djorg 2  & 3  & -159.9 & 0.5 & $\cdots$ & $\cdots$ & $\cdots$ \\ 
NGC 5634 & 10 & -17.9  & 0.4 & -45.1  & 6.6 & 27.2  \\ 
NGC 5694 & 9  & -131.5 & 0.6 & -140.3 & 0.8 & 8.8   \\ 
NGC 5946 & 10 & 114.2  & 0.4 & 128.4  & 1.8 & -14.2 \\ 
NGC 6256 & 10 & -103.4 & 0.5 & -101.4 & 1.9 & -2.0  \\ 
NGC 6284 & 7  & 20.7   & 0.4 & 27.5   & 1.7 & -6.8  \\ 
NGC 6316 & 3  & 92.2   & 0.6 & 71.4   & 8.9 & 20.8  \\ 
NGC 6352 & 12 & -123.7 & 0.3 & -137.0 & 1.1 & 13.3  \\ 
NGC 6355 & 7  & -210.3 & 0.4 & -176.9 & 7.1 & -33.4 \\ 
NGC 6366 & 14 & -118.3 & 0.3 & -122.2 & 0.5 & 3.9   \\ 
NGC 6401 & 7  & -115.4 & 0.3 & -65.0  & 8.6 & -50.4 \\ 
NGC 6426 & 7  & -220.0 & 1.0 & -162.0 & 23.0 & -58.0\\ 
NGC 6453 & 6  & -110.3 & 0.5 & -83.7  & 8.3 & -26.6 \\ 
NGC 6517 & 9  & -65.1  & 0.2 & -39.6  & 8.0 & -25.5 \\ 
NGC 6539 & 9  & 31.3   & 0.4 & 31.0   & 1.7 & 0.3   \\ 
NGC 6749 & 8  & -51.5  & 0.7 & -61.7  & 2.9 & 10.2  \\ 
NGC 6864 & 11 & -193.4 & 0.2 & -189.3 & 3.6 & -4.1  \\ 
Pal10   & 9  & -29.7  & 0.3 & -31.7  & 0.4 & 2.0   \\ 
Pal11   & 7  & -60.0  & 0.4 & -68.0  & 10.0 & 8.0  \\ 
Pal14   & 5  & 73.4   & 1.1 & 72.3   & 0.2 & 1.1   \\ 
Pal6    & 3  & 177.0  & 1.0 & 181.0  & 2.8 & -4.0  \\ 
Terzan 1 & 9  & 63.0   & 0.5 & 114.0  & 14.0 & -51.0\\ 
Terzan 2 & 3  & 144.6  & 0.8 & 109.0  & 15.0 & 35.6 \\ 
Terzan 8 & 7  & 142.5  & 0.4 & 130.0  & 8.0 & 12.5  \\ 
Terzan 9 & 6  & 71.4   & 0.4 & 59.0   & 10.0 & 12.4 \\ 
Ton 2    & 8  & -172.7 & 0.3 & -184.4 & 2.2 & 11.7  \\  
% &  &  &  &  &  &  \tabularnewline
\hline 
\end{tabular}
\end{centering} 

\tablefoot{
{Column 2 gives the number of member stars used to determine the mean radial velocity that is listed in Col.~3. Column 4 lists the standard {error} of the mean. The radial velocities in the literature (Cols.~5 and 6) are from the Harris catalogue, except for BH\,176, for which the source is \citet{sharina+2014}.  The difference $\Delta v_{\rm r}$ is given in the last column as FORS2$-$H10.}
}

\end{table}

Heliocentric radial velocities were measured for each spectrum using the IRAF \texttt{fxcor} task to perform cross-correlation between the target spectrum and a template spectrum with known radial velocity. This task provides observed radial velocities and heliocentric corrections using the celestial coordinates of the field and the observatory position.  The cross-correlation was done against a single synthetic spectrum for a typical metal-poor K giant star with $T_{\rm eff} = 4750$ K, {$\log g = 2.5$}, and $\rm [Fe/H] = -1.3$ covering the CaT region from 8450 \AA{} to 8750 \AA.  Using the selection criteria presented in \citet{saviane+2012}, candidate member stars were selected under the assumption that stars in a cluster have similar radial velocities and small dispersion of the order $\sim 10\,\rm km\,s^{-1}$ \citep{pryor+1993}. 
To define a first guess mean radial velocity for member selection, a 3$\sigma$ clipping iteration was used for all  stars in each cluster using the median and the first quartile as estimators.  After this pre-selection, the sum of the strengths of the $\lambda 8542$ and $\lambda 8662$ lines of the CaT was plotted  against $V-V_{{\rm HB}}$. In this plot, stars with similar metallicities are distributed along a line with fixed slope (see next section).   Data points deviating from the regression line defined by the median of candidate members were excluded by a new 3$\sigma$ clipping iteration. As an example, Fig.~\ref{selection} shows the selection process for the cluster NGC\,5946.

%\subsection{Radial velocities compared with Visible data}

\begin{figure}[t]
\centering
%  {\includegraphics[angle=0, width=1.0\columnwidth]{plots/03.2_RV_GC_comp_Vis_CaT.eps}}
%>  {\includegraphics[angle=0, width=1.0\columnwidth]{newplots/go_rv.eps}}
  {\includegraphics[angle=0, width=1.0\columnwidth]{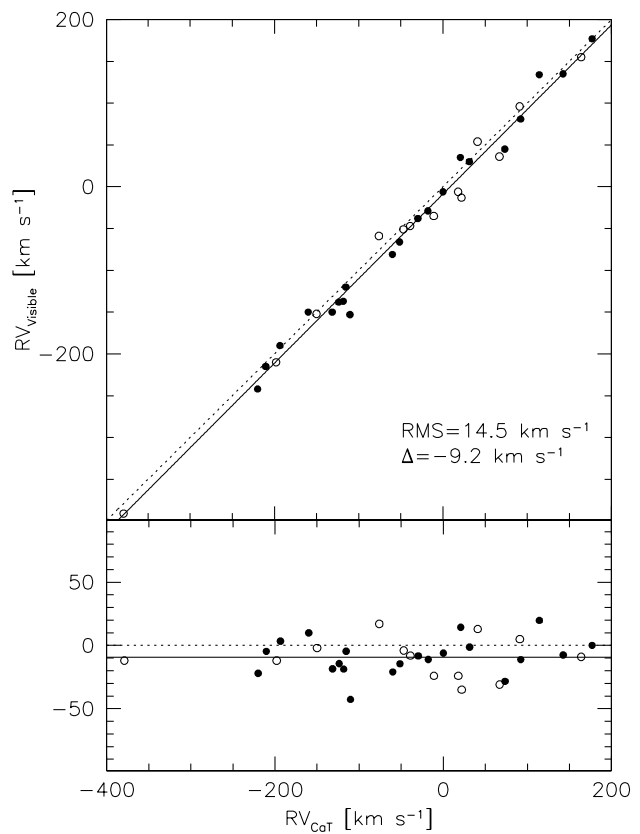}}
  \caption{Comparison between radial velocities for globular clusters analysed in \citet{saviane+2012} (open circles) and {this work (filled circles)}, and the measurements of \citet{dias+2016}. 
{In the upper panel the dotted line refers to the one-to-one relation, while the continuous line is a linear fit to the data. In the lower panel we plot the residuals $v_{\rm vis} -v_{\rm CaT} $.}
}
  \label{rv_comp_dias}
  \end{figure}

\begin{figure}
\centering
%>  {\includegraphics[angle=-90, width=1.0\columnwidth]{plots/03.3_vel_comp.eps}}
  {\includegraphics[angle=-90, width=1.0\columnwidth]{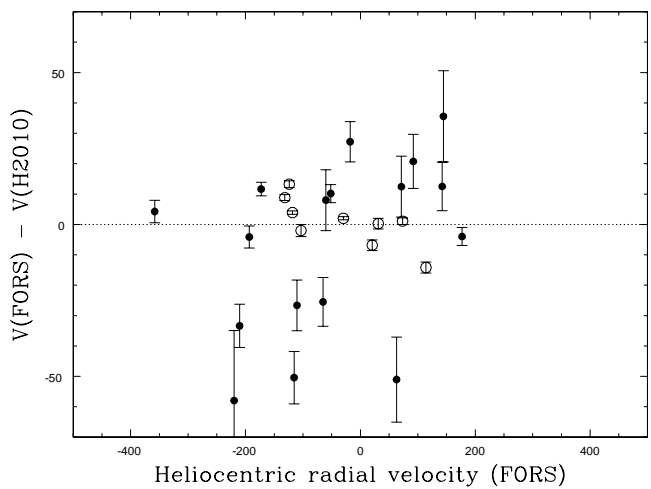}}
  \caption{Radial velocity comparison between our measurements and those listed by \citet{harris+1996}. The open circles are clusters with errors smaller than 2 km s$^{-1}$ in the Harris' catalogue, while the filled circles mark clusters with errors larger than 2 km s$^{-1}$.}
  \label{rv_comp}
  \end{figure}

The heliocentric radial velocities in this work were compared with those measured by our group 
{in the visual spectral range 4560 -- 5860 \AA\   \citep[][hereafter D16]{dias+2016}.}
The radial velocities of 34 GCs in common with the CaT sample (including the results of \citealt{saviane+2012}) are compared in Fig.~\ref{rv_comp_dias}.  The two independent sets of measurements show a small
  systematic offset $(v_{\rm CaT} - v_{\rm D16} = -9.2$ km s$^{-1})$. 
The radial velocity is more accurate in the CaT region {because of the larger number of sky lines used to correct the wavelength calibration.}
Our measurements are also compared in Fig.~\ref{rv_comp} with 
radial velocities published  for 26 clusters in the \citet{harris+1996} catalogue, 
{which are weighted averages of data collected from various published sources.} 
%
%% (BH\,176 and Djorg\,2 have no radial velocity data in the catalogue).  
The mean difference is $-4.5$ km s$^{-1}$ with a standard deviation 
$24$ km s$^{-1}$, 
a shift and dispersion that mostly originate from ten clusters with uncertainties larger than 8 km s$^{-1}$ in the catalogue. 
{In view of the homogeneous data quality and reduction in our work, we believe our values are systematically more accurate. Indeed, by picking only GCs with small uncertainties in the Harris' catalogue }
($< 2$ km s$^{-1}$), the mean and standard deviation of the difference are reduced to $-0.7$ km s$^{-1}$ and $10$ km s$^{-1}$ respectively. A compilation of radial velocity data for the GCs analysed here is given in Table \ref{tab:RV}, which, together with Table 2 in \citet{saviane+2012}, provides a homogeneous radial velocity data set for faint and reddened Galactic globular clusters.

%__________________________________________________________________
\section{Metallicity measurements}

%-----------------------------------------------
\subsection{Equivalent width measurements}\label{subsec:Equivalent-width-measurements}

\begin{figure}[ht]
\centering
{\includegraphics[width=1.0\columnwidth]{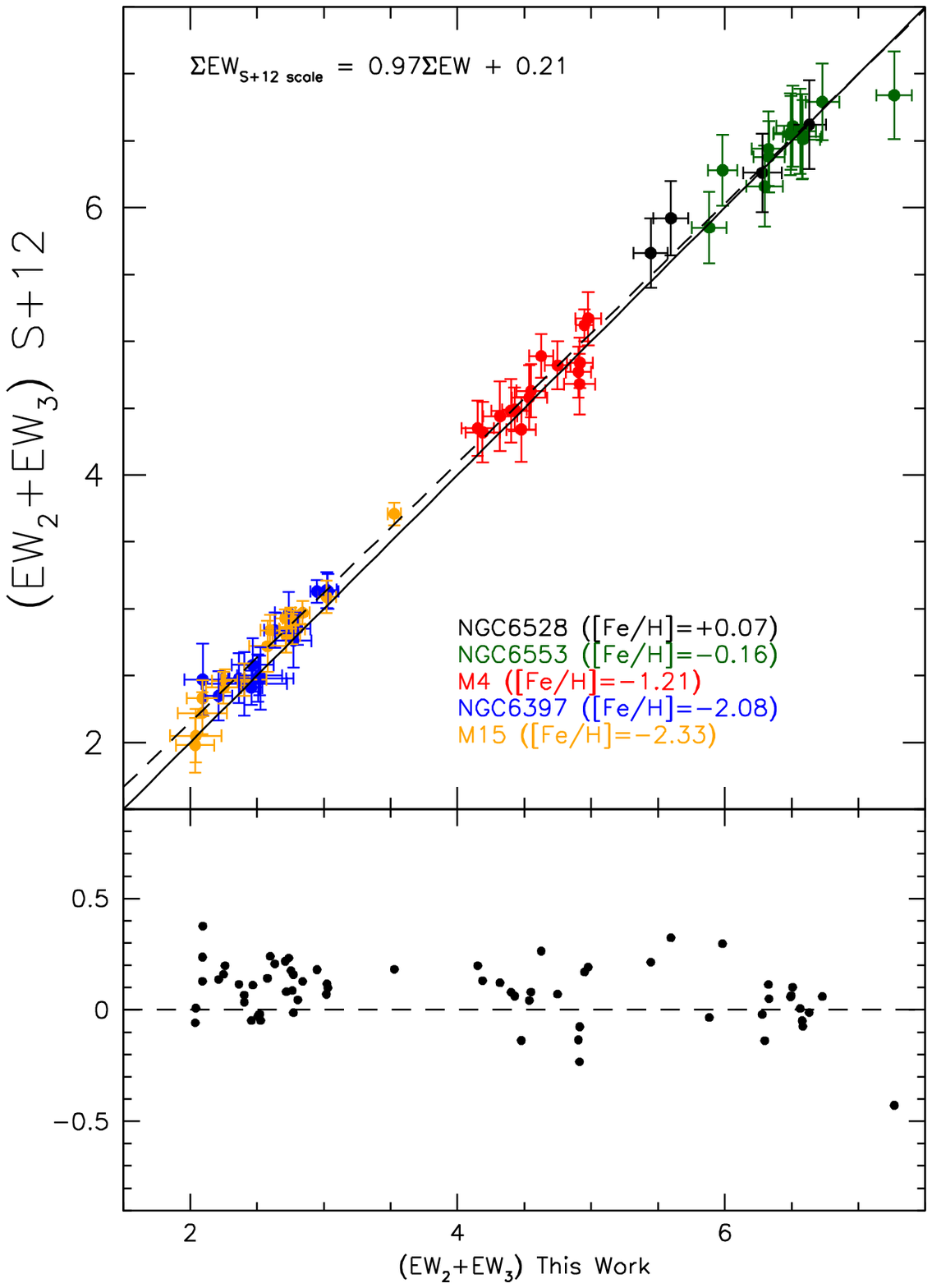}}
  \caption{Sum of EWs  of the two strongest CaT lines for stars in 5 template globular clusters measured by  \citet{saviane+2012} is plotted against our own  measurements. 
{The error bars were computed as in \citet{vasquez+2015}}. 
The one-to-one relation is shown as a black continuous line, whereas the dashed line is a linear fit to the data.  {The residuals from the one-to-one line are shown in the lower panel.} 
}
  \label{ew_comp}
  \end{figure}

\begin{figure}[ht]
\centering
  \resizebox{\hsize}{!}
 {\includegraphics[angle=-90]{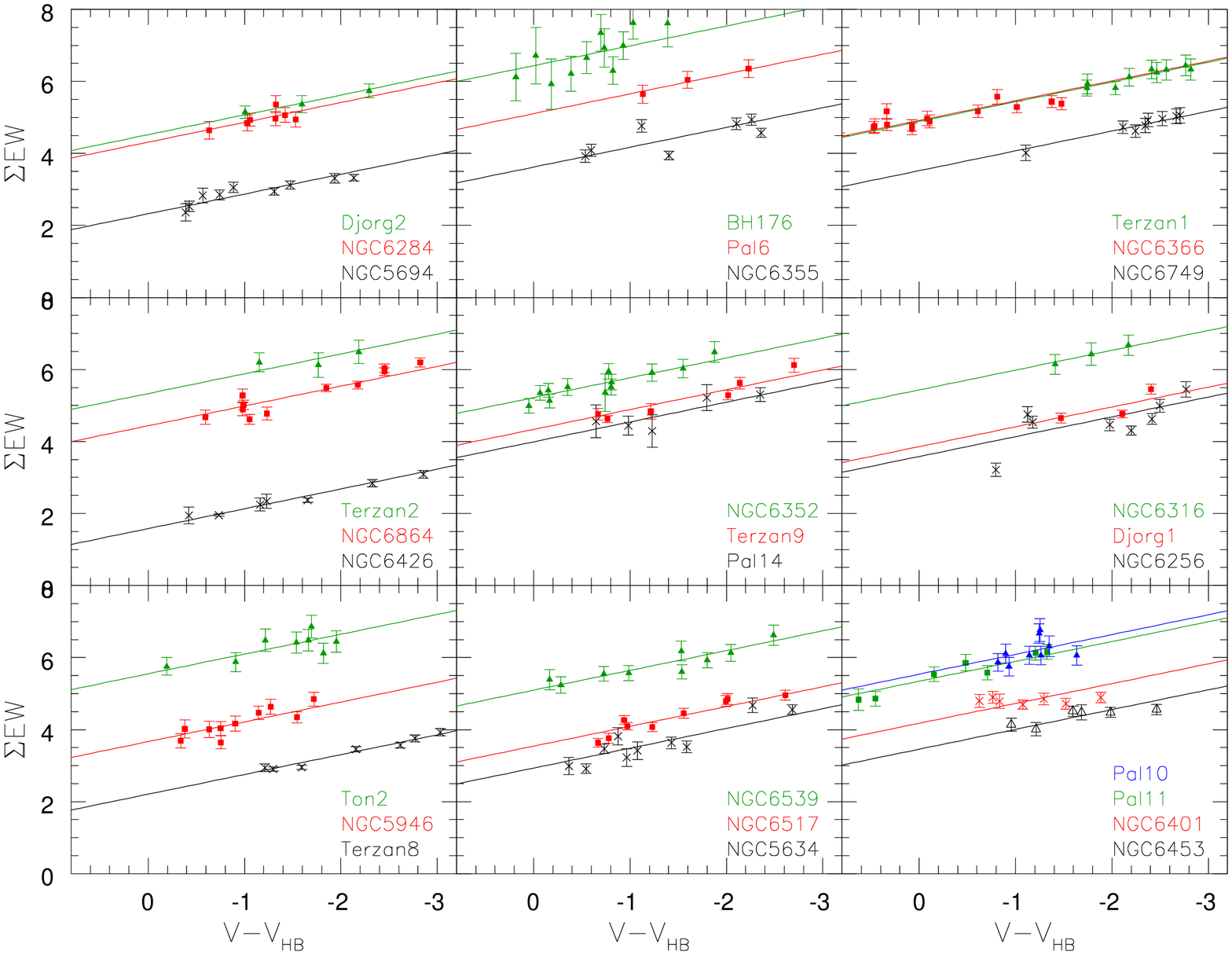}}
  \caption{Sum of the EWs of the two strongest CaT lines, $\Sigma_{\rm EW}$, is plotted against the magnitude difference from the horizontal branch ($V - V_{\rm HB}$) for the 28 programme clusters. Different colours are used to identify stars belonging to different clusters, together with {best fitting lines for the present data with a common fixed slope $\beta =-0.55$}.
}
  \label{clusters_all}
  \end{figure}

{The EWs were measured with the methods described in \citet{vasquez+2015}. As in Paper~I, }
 we used the sum of the EWs of the two strongest CaT lines ($\lambda8542$, $\lambda8662$) as a metallicity estimator, following the \ion{Ca}{ii} triplet method  of \citet{armandroff+1991}. 
Different functions have been tested in the literature to measure the EWs of the CaT lines, depending on the metallicity regime. For metal-poor stars ($\rm [Fe/H] \lesssim -0.7$ dex) a Gaussian function was used with excellent results \citep{armandroff+1991}, while a more general function (such as a Moffat function or the sum of a Gaussian and Lorentzian, G+L) is needed to fit the strong wings of the CaT lines observed in metal-rich stars \citep{rutledge+1997,cole+2004}. Following our previous work \citep{gullieus+2009,saviane+2012} we have adopted here a G+L profile fit. 
To measure the EWs, each spectrum was normalised with a low-order polynomial  
using the IRAF \texttt{continuum} task, and set to
the rest frame by correcting for the observed radial velocity. 
{The two strongest CaT lines were fitted by a G+L profile using a non-linear least squares fitting routine part of the \texttt{R} programming language.}
{Five clusters from the sample of \citet{saviane+2012} covering a wide metallicity range were re-reduced and analysed with our code to ensure that our EWs measurements are on the same scale as the template clusters used to define the metallicity calibration.}
Figure~\ref{ew_comp} shows the comparison between our EWs measurements and the line strengths measured by \citet{saviane+2012}  
(in both cases the sum of the two strongest lines) for the five calibration clusters.
The observed scatter is consistent with the internal errors of the EW measurements, {computed as in \citet{vasquez+2015}. }
The measurements show a small deviation from the unity relation, 
{which is more evident at low metallicity.}  A linear fit to this trend gives the relation: 
\(  \Sigma_{\rm EW} \textrm{(S12)} = 0.97\,\Sigma_{\rm EW} \textrm{(this work)} + 0.21 \), 
{with an rms about the fit of 0.13  \AA.}
This fit is shown in Fig. \ref{ew_comp} as a dashed black line.
{For internal consistency, all EWs in this work have been adjusted} to the measurement scale of \citet{saviane+2012} {by using this relation}.
In Table~\ref{tab:Program-Cluster-Data} we provide the coordinates, radial velocities, and the sum of the equivalent widths for the cluster member stars, both measured (``m'') and corrected (``c'') to the system of \citet{saviane+2012}.

%----------------------------------------------------------------------------------------------------------
\begin{table*}
\caption{Programme Cluster Data \label{tab:Program-Cluster-Data}}
%\begin{raggedright} %
\begin{tabular}{lll lrr ccl lll}
 &  &  &  &  & \tabularnewline
\hline \hline 
Name & Slit & RA (2000)  & Dec (2000)  & RV & eRV & ~~$\Delta V$  & 
$\Sigma_{\rm EW}$(c) & 
e$\Sigma_{\rm EW}$(c) & 
$\Sigma_{\rm EW}$(m) &
e$\Sigma_{\rm EW}$(m) 
\tabularnewline
& &  & & \multicolumn{2}{c}{(km s$^{-1}$)}  & (mag)  & 
(\AA{}) & & 
(\AA{}) &  
\tabularnewline
\hline
BH176 & 1\_668 & 15:39:12.4 & $-$50:2:18.9  & $-$2.30 & 1.53 & $-0.73$ & 6.93 & 0.53 & 6.92 & 0.22 \tabularnewline
BH176 & 1\_1578 & 15:39:12.6 & $-$50:1:59.0 & $-$5.42 & 1.43 & $-0.82$ & 6.30 & 0.39 & 6.28 & 0.16 \tabularnewline
BH176 & 1\_2513 & 15:39:10.4 & $-$50:1:38.6 & 4.63 & 3.62 &  $+$0.18 & 6.12 & 0.65 & 6.09 & 0.28 \tabularnewline
BH176 & 1\_3686 & 15:38:55.8 & $-$50:1:09.8 & 7.31 & 1.49 & $-1.03$ & 7.63 & 0.46 & 7.65 & 0.17 \tabularnewline
BH176 & 1\_4579 & 15:39:07.0 & $-$50:0:47.3 & 4.97 & 1.78 & $-0.39$ & 6.21 & 0.49 & 6.19 & 0.21 \tabularnewline
\hline \smallskip
\end{tabular}
%\end{raggedright}
\tablefoot{This table is available in its entirety in a machine-readable
form in the online journal. A few lines are shown here to illustrate the table content.
}
\end{table*}
%---------------------------------------------------------------------------------------------------------

%-------------------------------------
\subsection{Reduced equivalent widths}

According to the CaT method, the dependence on the stellar gravity and $T_{\rm eff}$  can be empirically corrected within a cluster by defining the reduced equivalent width $W^{\prime}$ as

\begin{equation}\label{w_prime}
W^{\prime}={\Sigma_{\rm EW} - \beta(V-V_{{\rm HB}})},
\end{equation}

where $V$ and $V_{{\rm HB}}$ are the visual magnitude of each RGB star and  the mean magnitude of the HB, respectively. The $V$ magnitude difference with respect to the HB is {not affected by errors in distance, reddening, or photometric zero-point}, while it is affected by internal differential reddening. 
In a plot of $\Sigma_{\rm EW}$   vs.   $(V-V_{\rm HB})$, {all measurements in a globular cluster follow a linear relation for member stars brighter than the HB (Fig.~\ref{clusters_all}).  The slope} is {approximately constant} in the metallicity range of globular clusters ($\rm -2.5<[Fe/H]<+0.0 $), although there is evidence that this assumption loses validity outside this range, especially for extremely metal poor stars \citep{starkenburg+2010,carrera+2013}. 
{The average slope calculated using our EW measurements is $\beta=-0.55$, only slightly different from the value found by \citet{saviane+2012}. }
%-- In this paper the average slope found by \citet{saviane+2012},   $\beta=-0.627$, is adopted.  
%
By removing the dependence on gravity and $T_{\rm eff}$, $W^{\prime}$  can be used to determine the metallicity of a given RGB star by using a suitable calibration.

Stellar magnitudes and the HB level have been measured using point spread function (PSF) photometry on the pre-imaging data, using Stetson's \noun{daophot/allstar} package \citep{stetson+1987,stetson+1994}. All the instrumental magnitudes were calibrated using colour terms and zero points provided by ESO as part of their routine quality control\footnote{see http://www.eso.org/observing/dfo/quality/FORS2/qc/photcoeff/\\
  photcoeffs\_fors2.html}.  As not all of the pre-imaging observations were taken in photometric conditions, the final magnitudes are only approximately on the $V$, $I$ standard system. However, this is not relevant to the metallicity calibration due to the differential dependence on $V-V_{{\rm HB}}$.

\begin{figure}[t]
  \centering{
 \includegraphics[width=1.0\columnwidth]{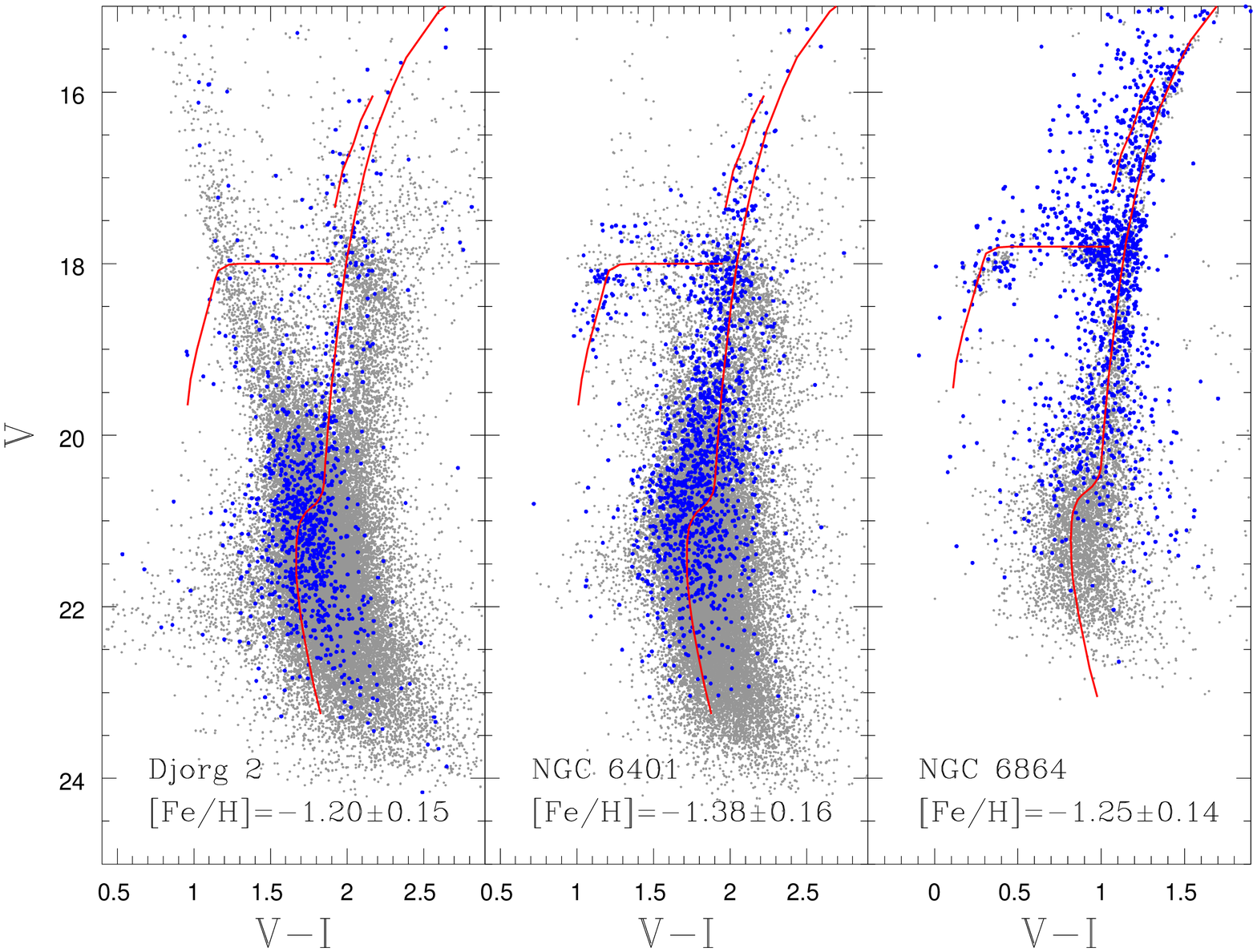}
  }  
  \caption{Optical CMDs of Djorg\,2, NGC\,6401 and NGC\,6864. The full photometric catalogues  (cluster + field)  obtained from our FORS2 pre-imaging are plotted in grey, while  blue dots are used for  the inner regions {($r<0.5$ arcmin)}. The fiducial loci published by \citet{catelan+2002} for NGC\,6864 are over-plotted as a red curve.}
  \label{djorg2_comp}
\end{figure}

To determine the HB level for each cluster, we used the CMD of NGC\,6864
to define a template HB. This cluster has a well defined blue HB along with a red HB, which makes the construction of a fiducial line easier.  As a first approximation we chose the fiducial line published by \citet{catelan+2002}, 
{shifted in colour and magnitude to match the observed CMD (see right panel of Fig.~\ref{djorg2_comp}).  
This HB template was used to match the CMDs using stars within the cluster half-light radius (given in the Harris catalogue). Examples are shown in Fig.~\ref{djorg2_comp}. } 
This selection minimises the field star 
contamination and makes it easier to recognise the HB morphology. As the
HB morphology changes with cluster metallicity, being blue and extended 
for metal poor clusters and limited to a red HB for
metal rich clusters, the matching criteria as well as the error on the HB
level differ from case to case.  For clusters with a red HB, the mean and
dispersion of red HB stars define $V_{\rm HB}$ and its
error (1$\sigma$). For metal poor clusters the mean level and its uncertainty (1$\sigma$ around the line) are defined by 
the stars close to the horizontal portion of the fiducial line. The measurement 
of $V_{\rm HB}$ is also affected by differential reddening, which is
difficult to correct without a proper reddening map for the cluster. 
To take into account the uncertainty introduced by this effect,
no selection was applied to the stars used to measure $V_{\rm HB}$. 
The variable reddening increases the 1$\sigma$ error on 
$V_{\rm HB}$. The most uncertain value was derived for
Terzan\,1 and \object{Terzan\,2} with $\Delta V_{\rm HB}\sim0.35$ mag, which leads 
to an uncertainty of $\sim0.15$ dex in metallicity. For the
rest of the sample the errors on $V_{\rm HB}$ are typically $\sim 0.1$
mag. {The $V-V_{{\rm HB}}$ values of member stars are given in Table~\ref{tab:Program-Cluster-Data}.}

%------------------------------------------------------------------------------------------------------------
\subsection{Metallicity calibration}
\label{calibEW}

\begin{figure*}
%\begin{tabular}{cc}
\centering
\includegraphics[height=\textwidth,angle=-90]{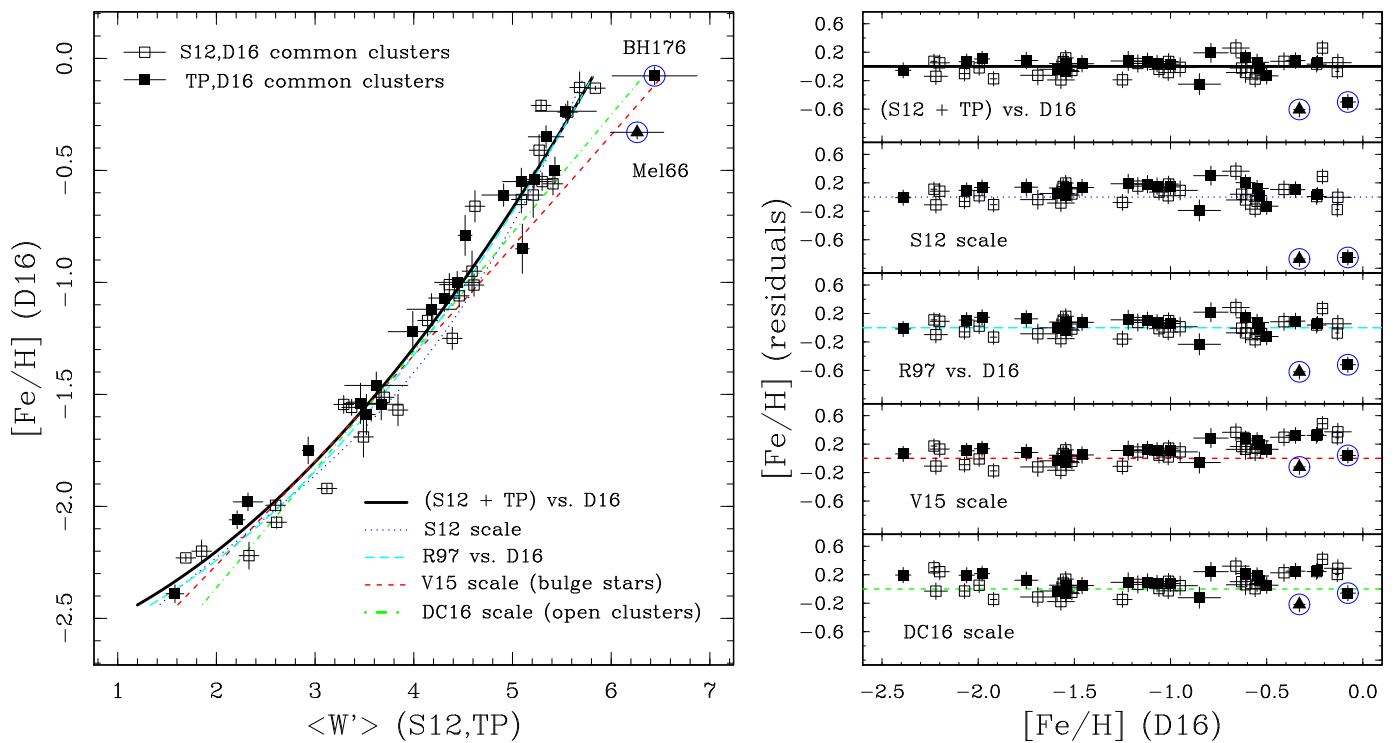}
\caption{
Comparison  between the metallicities of GGCs derived from the optical spectral range \citep{dias+2016} and the mean reduced equivalent widths 
$\langle W^\prime \rangle$\  for the clusters in common with data in this paper  {(filled squares)} and the S12 sample {(open squares)}. The peculiar cluster BH\,176 and the open cluster Mel\,66 are marked by open circles. 
Different CaT calibrations are shown: 
the best fit to the merged FORS2 sample as a black solid line, the \citet{saviane+2012} relation as a blue dotted line, and the \citet{rutledge+1997} calibration (converted to our $W^\prime$ scale) as a cyan dashed line. These relations based on globular cluster data share a common curved trend. 
In addition, the \citet{vasquez+2015} fit which includes bulge stars on the metal-rich side is shown as a red dashed line, and the calibration of \citet{dacosta+2016} {based on stars in globular and open clusters} is represented by the green dot-dashed line.  In the right panels we plot the [Fe/H] residuals (this work $-$ D16) for five CaT calibrations. See discussion in the text.}
\label{fig:Metallicity-values-from}
\end{figure*}

%------- QUADRATIC CALIBRATIONS

In \citet{saviane+2012} the reduced equivalent widths were converted
into ${\rm [Fe/H]}$ values using the empirical calibration 
\begin{equation}
{\rm [Fe/H]_{S12}}=0.0178W^{\prime\,3}-0.114W^{\prime\,2}+0.599W^{\prime}-3.113,\label{eq:s12}
\end{equation}
based on a set of template globular clusters with well-established abundances from high resolution spectroscopy. This relation has an r.m.s. dispersion around the fit of 0.13 dex and is defined in the ${\rm [Fe/H]}$ range from $-2.33\,{\rm dex}$ to $+0.07\,{\rm dex}$.
A cubic relation between $W^{\prime}$ and ${\rm [Fe/H]}$ was needed to fit the most metal-rich and metal-poor globular clusters in the calibration sample. 
A {slight curvature} in the CaT calibration appears to be common to several published calibrations based on globular clusters \citep[e.g., ][]{rutledge+1997,carretta+2009,carrera+2013}. 
As long as the calibration templates and the target clusters share the same chemical
evolution history and are measured in the same way (as in our case), 
the shape of the empirical calibration is not critical and there is no theoretical reason for it to be linear. 
The main caveat is represented here by the paucity of globular cluster calibrators at high metallicity: 
indeed, Eq.~(\ref{eq:s12}) was based on just three GCs more metal-rich that {[}Fe/H{]}=$-0.7$.

On the other hand, linear calibrations have been obtained by extending the fit using metal-rich stars chosen outside the globular cluster population \citep[e.g., ][ and references therein]{cole+2004,carrera+2013}. 
The most recent examples are those of  \citet{vasquez+2015} and \citet{dacosta+2016}.
\citet{vasquez+2015} proposed a new metallicity calibration on the C09 scale, using red giant and red clump stars in the Galactic bulge to define the metal-rich end of the calibration, and globular clusters giants from \citet{warren+2009} at the metal-poor end.  Each sample was analysed independently, finding linear calibrations between $W^{\prime}$ and metallicity with a small difference in slope ($< 7$\%).  To take into account this small difference, a quadratic polynomial was used to link the fits to the bulge stars and GCs:
\begin{equation}
{\rm [Fe/H]_{V15}}=0.006W^{\prime\,2}+0.432W^{\prime}-3.150,\label{eq:v14}
\end{equation}
with an r.m.s. spread of $0.197$ dex. The curvature of this relation is very modest so it  can be regarded as nearly linear. 
{\citet{dacosta+2016} used a combination of Galactic globular clusters and open clusters} to establish a linear relation between $W^{\prime}$
and ${\rm [Fe/H]}$ in the metallicity range ${\rm -2.4<[Fe/H]<+0.1}$.
The fit  is: 
\begin{equation}
{\rm [Fe/H]_{DC16}} = (0.528\pm0.017) W^{\prime} - (3.420\pm0.077), \label{eq:dc16}
\end{equation}
with an r.m.s. of the residuals of $0.06$ dex.  {This relation is consistent with that of \citet{saviane+2012} up to the metallicity of 47\,Tuc.}

All these $\mbox{[Fe/H]} -  \langle W^{\prime} \rangle$ calibrations are shown in Fig.~\ref{fig:Metallicity-values-from} (left panel), along with the \citet{rutledge+1997} calibration corrected to the $W^{\prime}$  definition of \citet{saviane+2012}.  The relations are consistent for metallicities up to [Fe/H]$\sim -0.7$, while they diverge for metal-rich clusters. 
In Fig.~\ref{fig:Metallicity-values-from} they are over-plotted on a sub-set of our catalogue in common with \citet{dias+2015,dias+2016}\footnote{48 of our 56 clusters have been observed in the framework of our GGC survey using the same slit masks in the optical and near-infrared wavelength ranges.}.
In the parallel work of Dias et al., metallicities were obtained from spectral fitting of medium-resolution spectra in the interval 4560 -- 5860~\AA, along  with estimates of the stellar parameters and  alpha-element abundances.  Since the [Fe/H] values of \citet{dias+2016} are in excellent agreement with high-resolution spectroscopic metallicities in the range -2.5 $<$ [Fe/H] $<$ 0.0  (see discussion in D16), they provide an independent CaT calibration: 
\begin{equation}
{\rm [Fe/H]_{D16}} = (0.055\pm0.012) W^{\prime\,2} + (0.13\pm0.10) W^{\prime} - (2.68\pm0.17)
\label{V17S12scale}
\end{equation}
%}
with r.m.s. $= 0.11$~dex and coefficient of determination $r^2=0.97$. 
The new relation is shown in Fig.~\ref{fig:Metallicity-values-from} (left panel) as a continuous line. The fit appears to confirm the S12 calibration in the validity range for globular clusters.  However, it is now based on 17 clusters more metal-rich than {[}Fe/H{]}=$-0.7$, giving higher weight to the non-linear behaviour found for globular clusters by S12 and other authors.  

%----------------------------------------------------------

The right panels of Fig.~\ref{fig:Metallicity-values-from} show the residuals between the [Fe/H] values computed using different CaT calibrations and the metallicities in D16 for all clusters in common. 
The CaT metallicities are the average of  individual metallicities computed for member stars. 
{The error bars represent the uncertainties in our determinations which are a combination of the statistical uncertainty in the
EWs, the calibration uncertainty (taken as the rms dispersion about the fit) and, where necessary, inclusion of allowance
for significant differential reddening that can affect the individual  $V_{\rm HB}$  values (see S12 for details).}
It is apparent from Fig.~\ref{fig:Metallicity-values-from} that non-linear calibrations  provide a better fit for globular clusters, with the exception of the metal-rich cluster BH\,176 which has anomalously large  $W^{\prime}$. 
The position of BH\,176 in the {[}Fe/H{]} vs. $\langle W^{\prime} \rangle$ plot is more similar to the open cluster Melotte\,66 than to other metal-rich GCs. 
The $\langle W^{\prime} \rangle$ of Melotte\,66 was remeasured with our code using spectra from \citet{dacosta+2016}. Our $\langle W^{\prime} \rangle$ value given in Fig.~\ref{fig:Metallicity-values-from}, although larger than the original measurement of \citet{dacosta+2016} lying pretty much exactly on the author's calibration line, confirms that the $W^{\prime}$ values measured for  stars in this open cluster (as well as in BH\,176) are larger than those obtained in GCs of similar metallicity. 
%
%%Thus, systematic effects related to the different codes do not provide a plausible explanation.
%
At present, a full understanding of the calibration of the CaT line strengths for metal-rich stars  calls for further investigation and theoretical modelling of behaviour of these lines in giant stars as a function of  
{the stellar parameters $\log g$, $T_{\rm eff}$, [Fe/H], and [$\alpha$/Fe] or, equivalently, the population properties age and metallicity.}

The metallicities of the 28 globular clusters measured in this paper are therefore listed in Table~\ref{tab:cluster_paper2} using different calibrations, although we maintain the S12 metallicity scale as our reference calibration. 
{The reduced EWs listed in these tables are given on the measurement system of \citet{saviane+2012}}  
and the metallicities are calculated on the scales of 
\citet[][Eq. \ref{V17S12scale}]{dias+2016}, \citet[][Eq. \ref{eq:s12}]{saviane+2012},
and \citet[][Eq.~\ref{eq:v14}]{vasquez+2015}.  
The errors are computed as in S12 and Fig.~\ref{fig:Metallicity-values-from}. 
{We note that, in some cases, the metallicities are based on a few member stars only and should be regarded as more dubious than suggested by the internal error.} 
In the case of the V15 calibration the errors are dominated by a large calibration uncertainty of 0.197 dex.
{Further, the listed $W^{\prime}$ can be used to calculate the metallicities on the scale of \citet[][Eq.~\ref{eq:dc16}]{dacosta+2016}.}
For the sake of completeness, the clusters in \citet{saviane+2012} are listed in Table~\ref{tab:cluster_paper1} using the same calibrations.
{The cluster metallicities from the compilations of \citet{dias+2016eso}, \citet{harris+1996} and C09 are also given for comparison.}

%--------------------------------------------------------------------------------------------------------------------
\begin{table*}
\caption{Metallicities of programme clusters using the D16, S12, and V15 calibrations, compared with [Fe/H] from the literature
\label{tab:cluster_paper2}}
\small
%\centering 
\begin{tabular}{lccccc|lcccc}
\hline 
\noalign{\smallskip}
Cluster  & $N$  & $\left\langle W'\right\rangle $ & ${\rm [Fe/H]}_{\rm CaT}$ & ${\rm [Fe/H]}_{\rm CaT}$ & ${\rm [Fe/H]}_{\rm CaT}$ 
& $~~~{\rm [Fe/H]_{D16}}$  & ${\rm [Fe/H]_{H10}}$  & $w$  & ${\rm [Fe/H]_{C09}}$   & $\rho$ \tabularnewline
 &   & & D16 cal. & S12 cal. & V15 cal. &  &   &   & \tabularnewline
%  &   & & D16 scale & S12 scale & V15 scale & D16 avg  &  \multicolumn{2}{c}{H10 avg}  &   C09 avg &\tabularnewline
%%% (1)  & (2)  & (3)  & (4)  & (5)  & (6)  & (7)  & (8)  & (9)  & (10) & (11) \tabularnewline
\hline   
\noalign{\smallskip} %\input{ivotab4_d16a.tex}
BH176  &  11 & $6.44\pm 0.43$ & $\cdots$ & $\cdots$ & $-0.12\pm 0.22$ & $-0.08\pm 0.04$ & $0.00$ & $1$ & $\cdots$ & $0.7$ \tabularnewline
Djorg1  &  3 & $3.86\pm 0.26$ & $-1.36\pm 0.14$ & $-1.48\pm 0.15$ & $-1.39\pm 0.21$ & $-1.54\pm 0.13\tablefootmark{a}$ & $-1.51$ & $1$ & $\cdots$ & $1.5$ \tabularnewline
Djorg2  &  3 & $4.52\pm 0.07$ & $-0.97\pm 0.13$ & $-1.09\pm 0.15$ & $-1.07\pm 0.21$ & $-0.79\pm 0.09$ & $-0.65$ & $1$ & $\cdots$ & $0.4$ \tabularnewline
NGC5634  &  10 & $2.93\pm 0.28$ & $-1.83\pm 0.12$ & $-1.89\pm 0.13$ & $-1.83\pm 0.20$ & $-1.75\pm 0.06$ & $-1.88$ & $2$ & $-1.93\pm 0.09$ & $1.3$ \tabularnewline
NGC5694  &  9 & $2.32\pm 0.15$ & $-2.08\pm 0.11$ & $-2.11\pm 0.13$ & $-2.12\pm 0.20$ & $-1.98\pm 0.06$ & $-1.98$ & $5$ & $-2.02\pm 0.07$ & $1.0$ \tabularnewline
NGC5946  &  10 & $3.67\pm 0.22$ & $-1.46\pm 0.13$ & $-1.57\pm 0.14$ & $-1.48\pm 0.21$ & $-1.54\pm 0.07$ & $-1.29$ & $4$ & $-1.29\pm 0.14$ & $0.9$ \tabularnewline
NGC6256  &  10 & $3.58\pm 0.44$ & $-1.51\pm 0.14$ & $-1.61\pm 0.14$ & $-1.53\pm 0.21$ & $-1.05\pm 0.13\tablefootmark{a}$ & $-1.02$ & $2$ & $-0.62\pm 0.09$ & $2.2$ \tabularnewline
NGC6284  &  7 & $4.31\pm 0.16$ & $-1.10\pm 0.12$ & $-1.22\pm 0.14$ & $-1.18\pm 0.20$ & $-1.07\pm 0.06$ & $-1.26$ & $2$ & $-1.31\pm 0.09$ & $0.7$ \tabularnewline
NGC6316  &  3 & $5.43\pm 0.06$ & $-0.35\pm 0.14$ & $-0.37\pm 0.17$ & $-0.63\pm 0.21$ & $-0.50\pm 0.06$ & $-0.45$ & $5$ & $-0.36\pm 0.14$ & $0.1$ \tabularnewline
NGC6352  &  12 & $5.22\pm 0.18$ & $-0.50\pm 0.12$ & $-0.56\pm 0.14$ & $-0.73\pm 0.20$ & $-0.54\pm 0.04$ & $-0.64$ & $5$ & $-0.62\pm 0.05$ & $0.7$ \tabularnewline
NGC6355  &  7 & $3.62\pm 0.32$ & $-1.49\pm 0.14$ & $-1.59\pm 0.15$ & $-1.51\pm 0.21$ & $-1.46\pm 0.06$ & $-1.37$ & $4$ & $-1.33\pm 0.14$ & $1.6$ \tabularnewline
NGC6366  &  14 & $4.91\pm 0.22$ & $-0.72\pm 0.12$ & $-0.81\pm 0.14$ & $-0.88\pm 0.20$ & $-0.61\pm 0.05$ & $-0.59$ & $3$ & $-0.59\pm 0.08$ & $1.3$ \tabularnewline
NGC6401  &  7 & $4.18\pm 0.25$ & $-1.18\pm 0.14$ & $-1.30\pm 0.15$ & $-1.24\pm 0.21$ & $-1.12\pm 0.07$ & $-1.02$ & $5$ & $-1.01\pm 0.14$ & $1.3$ \tabularnewline
NGC6426  &  7 & $1.58\pm 0.09$ & $-2.34\pm 0.11$ & $-2.38\pm 0.13$ & $-2.45\pm 0.20$ & $-2.39\pm 0.04$ & $-2.15$ & $3$ & $\cdots$ & $0.8$ \tabularnewline
NGC6453  &  6 & $3.46\pm 0.17$ & $-1.57\pm 0.12$ & $-1.67\pm 0.13$ & $-1.58\pm 0.20$ & $-1.54\pm 0.09$ & $-1.50$ & $4$ & $-1.48\pm 0.14$ & $1.0$ \tabularnewline
NGC6517  &  9 & $3.54\pm 0.18$ & $-1.53\pm 0.12$ & $-1.63\pm 0.13$ & $-1.55\pm 0.20$ & $-1.26\pm 0.13\tablefootmark{a}$ & $-1.23$ & $4$ & $-1.24\pm 0.14$ & $1.1$ \tabularnewline
NGC6539  &  9 & $5.09\pm 0.19$ & $-0.59\pm 0.12$ & $-0.67\pm 0.14$ & $-0.80\pm 0.20$ & $-0.55\pm 0.06$ & $-0.63$ & $5$ & $-0.53\pm 0.14$ & $0.8$ \tabularnewline
NGC6749  &  8 & $3.52\pm 0.09$ & $-1.54\pm 0.11$ & $-1.64\pm 0.13$ & $-1.56\pm 0.20$ & $-1.59\pm 0.09$ & $-1.60$ & $1$ & $-1.62\pm 0.09$ & $0.3$ \tabularnewline
NGC6864  &  11 & $4.44\pm 0.23$ & $-1.02\pm 0.12$ & $-1.14\pm 0.14$ & $-1.11\pm 0.20$ & $-1.00\pm 0.05$ & $-1.29$ & $3$ & $-1.29\pm 0.14$ & $1.3$ \tabularnewline
Pal10  &  9 & $5.54\pm 0.31$ & $-0.27\pm 0.16$ & $-0.27\pm 0.19$ & $-0.57\pm 0.21$ & $-0.24\pm 0.03$ & $-0.10$ & $1$ & $\cdots$ & $1.0$ \tabularnewline
Pal11  &  7 & $5.34\pm 0.18$ & $-0.42\pm 0.13$ & $-0.45\pm 0.15$ & $-0.67\pm 0.20$ & $-0.35\pm 0.05$ & $-0.40$ & $4$ & $-0.45\pm 0.08$ & $0.7$ \tabularnewline
Pal14  &  5 & $3.99\pm 0.25$ & $-1.29\pm 0.13$ & $-1.41\pm 0.14$ & $-1.33\pm 0.21$ & $-1.22\pm 0.09$ & $-1.62$ & $4$ & $-1.63\pm 0.08$ & $0.7$ \tabularnewline
Pal6  &  3 & $5.10\pm 0.07$ & $-0.59\pm 0.14$ & $-0.66\pm 0.16$ & $-0.79\pm 0.21$ & $-0.85\pm 0.11$ & $-0.91$ & $3$ & $-1.06\pm 0.09$ & $0.2$ \tabularnewline
Terzan1  &  9 & $4.88\pm 0.10$ & $-0.74\pm 0.18$ & $-0.84\pm 0.21$ & $-0.90\pm 0.23$ & $-1.06\pm 0.13\tablefootmark{a}$ & $-1.03$ & $3$ & $-1.29\pm 0.09$ & $0.3$ \tabularnewline
Terzan2  &  3 & $5.33\pm 0.21$ & $-0.42\pm 0.18$ & $-0.46\pm 0.22$ & $-0.68\pm 0.22$ & $-0.72\pm 0.13\tablefootmark{a}$ & $-0.69$ & $3$ & $-0.29\pm 0.09$ & $0.6$ \tabularnewline
Terzan8  &  7 & $2.21\pm 0.08$ & $-2.12\pm 0.11$ & $-2.15\pm 0.13$ & $-2.17\pm 0.20$ & $-2.06\pm 0.04$ & $-2.16$ & $2$ & $\cdots$ & $0.8$ \tabularnewline
Terzan9  &  6 & $4.34\pm 0.19$ & $-1.08\pm 0.14$ & $-1.21\pm 0.15$ & $-1.16\pm 0.21$ & $-1.08\pm 0.13\tablefootmark{a}$ & $-1.05$ & $2$ & $-2.07\pm 0.09$ & $0.7$ \tabularnewline
Ton2  &  8 & $5.55\pm 0.26$ & $-0.26\pm 0.15$ & $-0.26\pm 0.19$ & $-0.57\pm 0.21$ & $-0.73\pm 0.13\tablefootmark{a}$ & $-0.70$ & $1$ & $\cdots$ & $0.8$ \tabularnewline
\hline 
\end{tabular}
\tablefoottext{$a$}{metallicities from \citet{dias+2016eso}}
\tablefoot{Columns: (1) Cluster ID. (2) Number of cluster member stars analysed. (3) Average reduced EW of clusters in our sample  on the measurement scale of \citet{saviane+2012}, as discussed in Sect.~\ref{subsec:Equivalent-width-measurements}.
{(4) Metallicity calibrated on the D16 scale} using Eq. \ref{V17S12scale}, and total uncertainty. (5,6) The same for the S12 and V15 scales, using Eq.~\ref{eq:s12} and Eq.~\ref{eq:v14}. (7) {Average metallicity from \citet{dias+2016} if present, or the homogeneous compilation by \citet{dias+2016eso} on the metallicity scale of \citet{dias+2016}.} (8,9) Metallicity and weight from the Harris catalogue. (10) Average metallicity from literature compiled by \citet{carretta+2009}. (11) Ratio of the rms dispersion around the linear fit with fixed slope, to the mean measurement error.}
\end{table*}

%----------------------------------------------------------------------------------------------------------------------

\begin{table*}
%\caption{As Table~\ref{tab:cluster_paper2}, for the clusters in  \citet{saviane+2012}
\caption{Metallicities of globular clusters in \citet{saviane+2012}
\label{tab:cluster_paper1}}
\small
\centering %
\begin{tabular}{lccccc|lcccc}
\hline 
\noalign{\smallskip}
Cluster  & $N$  & $\left\langle W'\right\rangle $ & ${\rm [Fe/H]}_{\rm CaT}$ & ${\rm [Fe/H]}_{\rm CaT}$ & ${\rm [Fe/H]}_{\rm CaT}$ 
& $~~~{\rm [Fe/H]_{D16}}$  & ${\rm [Fe/H]_{H10}}$  & $w$  & ${\rm [Fe/H]_{C09}}$   & $\rho$ \tabularnewline
 &   & & D16 cal. & S12 cal. & V15 cal. &  &   &   & \tabularnewline
%  &   & & D16 scale & S12 scale & V15 scale & D16 avg  &  \multicolumn{2}{c}{H10 avg}  &   C09 avg &\tabularnewline
%%% (1)  & (2)  & (3)  & (4)  & (5)  & (6)  & (7)  & (8)  & (9)  & (10) & (11) \tabularnewline
\hline 
\noalign{\smallskip}  %\input{ivotab5_d16a.tex}
Pyxis  &  5 & $3.90\pm 0.24$ & $-1.34\pm 0.13$ & $-1.45\pm 0.14$ & $-1.37\pm 0.20$ & $-1.40\pm 0.13\tablefootmark{a}$ & $-1.20$ & $3$ & $\cdots$ & $0.8$ \tabularnewline
NGC2808  &  17 & $4.46\pm 0.20$ & $-1.01\pm 0.11$ & $-1.13\pm 0.13$ & $-1.10\pm 0.20$ & $-1.06\pm 0.05$ & $-1.14$ & $4$ & $-1.18\pm 0.04$ & $1.0$ \tabularnewline
Rup106  &  9 & $3.29\pm 0.25$ & $-1.66\pm 0.12$ & $-1.74\pm 0.13$ & $-1.66\pm 0.20$ & $-1.54\pm 0.04$ & $-1.68$ & $6$ & $-1.78\pm 0.08$ & $1.9$ \tabularnewline
NGC5824  &  17 & $2.60\pm 0.30$ & $-1.97\pm 0.11$ & $-2.01\pm 0.13$ & $-1.99\pm 0.20$ & $-1.99\pm 0.03$ & $-1.91$ & $4$ & $-1.94\pm 0.14$ & $2.2$ \tabularnewline
Lynga7  &  8 & $5.21\pm 0.11$ & $-0.51\pm 0.13$ & $-0.57\pm 0.15$ & $-0.74\pm 0.20$ & $-0.61\pm 0.10$ & $-1.01$ & $3$ & $\cdots$ & $0.3$ \tabularnewline
NGC6139  &  15 & $3.54\pm 0.15$ & $-1.53\pm 0.11$ & $-1.63\pm 0.13$ & $-1.55\pm 0.20$ & $-1.58\pm 0.07\tablefootmark{a}$ & $-1.65$ & $2$ & $-1.71\pm 0.09$ & $1.0$ \tabularnewline
Terzan3  &  10 & $4.54\pm 0.12$ & $-0.96\pm 0.13$ & $-1.08\pm 0.14$ & $-1.07\pm 0.20$ & $-1.01\pm 0.08\tablefootmark{a}$ & $-0.74$ & $2$ & $\cdots$ & $0.4$ \tabularnewline
NGC6325  &  10 & $3.99\pm 0.20$ & $-1.29\pm 0.12$ & $-1.41\pm 0.13$ & $-1.33\pm 0.20$ & $-1.38\pm 0.10\tablefootmark{a}$ & $-1.25$ & $4$ & $-1.37\pm 0.14$ & $1.2$ \tabularnewline
NGC6356  &  11 & $5.30\pm 0.15$ & $-0.45\pm 0.11$ & $-0.49\pm 0.13$ & $-0.69\pm 0.20$ & $-0.55\pm 0.04$ & $-0.40$ & $5$ & $-0.35\pm 0.14$ & $0.5$ \tabularnewline
HP1  &  8 & $4.14\pm 0.30$ & $-1.20\pm 0.13$ & $-1.32\pm 0.14$ & $-1.26\pm 0.20$ & $-1.17\pm 0.07$ & $-1.00$ & $4$ & $-1.57\pm 0.09$ & $1.3$ \tabularnewline
NGC6380  &  8 & $5.31\pm 0.18$ & $-0.44\pm 0.14$ & $-0.48\pm 0.16$ & $-0.69\pm 0.20$ & $-0.47\pm 0.14\tablefootmark{a}$ & $-0.75$ & $2$ & $-0.40\pm 0.09$ & $0.7$ \tabularnewline
NGC6440  &  8 & $5.56\pm 0.18$ & $-0.26\pm 0.13$ & $-0.25\pm 0.16$ & $-0.56\pm 0.20$ & $-0.24\pm 0.05$ & $-0.36$ & $6$ & $-0.20\pm 0.14$ & $0.6$ \tabularnewline
NGC6441  &  7 & $5.27\pm 0.17$ & $-0.47\pm 0.14$ & $-0.52\pm 0.16$ & $-0.71\pm 0.20$ & $-0.41\pm 0.07$ & $-0.46$ & $6$ & $-0.44\pm 0.07$ & $0.6$ \tabularnewline
NGC6558  &  4 & $4.61\pm 0.15$ & $-0.91\pm 0.12$ & $-1.03\pm 0.14$ & $-1.03\pm 0.20$ & $-1.01\pm 0.05$ & $-1.32$ & $5$ & $-1.37\pm 0.14$ & $0.9$ \tabularnewline
Pal7  &  14 & $5.41\pm 0.12$ & $-0.37\pm 0.12$ & $-0.39\pm 0.14$ & $-0.64\pm 0.20$ & $-0.56\pm 0.05$ & $-0.75$ & $1$ & $-0.65\pm 0.09$ & $0.7$ \tabularnewline
NGC6569  &  7 & $4.62\pm 0.19$ & $-0.91\pm 0.14$ & $-1.02\pm 0.16$ & $-1.03\pm 0.21$ & $-0.66\pm 0.07$ & $-0.76$ & $4$ & $-0.72\pm 0.14$ & $0.8$ \tabularnewline
NGC6656  &  41 & $3.12\pm 0.30$ & $-1.74\pm 0.11$ & $-1.81\pm 0.13$ & $-1.74\pm 0.20$ & $-1.92\pm 0.02$ & $-1.70$ & $8$ & $-1.70\pm 0.08$ & $2.3$ \tabularnewline
NGC6715  &  15 & $3.30\pm 0.25$ & $-1.65\pm 0.11$ & $-1.74\pm 0.13$ & $-1.66\pm 0.20$ & $-1.31\pm 0.04\tablefootmark{a}$ & $-1.49$ & $7$ & $-1.44\pm 0.07$ & $2.5$ \tabularnewline
Terzan7  &  7 & $5.62\pm 0.12$ & $-0.21\pm 0.12$ & $-0.19\pm 0.13$ & $-0.53\pm 0.20$ & $-0.23\pm 0.10\tablefootmark{a}$ & $-0.32$ & $6$ & $-0.12\pm 0.08$ & $0.4$ \tabularnewline
NGC7006  &  18 & $3.49\pm 0.25$ & $-1.56\pm 0.11$ & $-1.65\pm 0.13$ & $-1.57\pm 0.20$ & $-1.69\pm 0.09$ & $-1.52$ & $6$ & $-1.46\pm 0.06$ & $1.2$ \tabularnewline
\hline 
\end{tabular}
\tablefoot{Columns as in Table~\ref{tab:cluster_paper2}. The average reduced EW in Col.3 are from \citet{saviane+2012}. 
}
\end{table*}
%----------------------------------------------------------------------------------------------------------------------

%__________________________________________________________________
%__________________________________________________________________
\section{Individual cluster results}

Some clusters present special characteristics in terms of possible metallicity spread, differential reddening effects and differences in radial velocity and [Fe/H] with respect to the literature. As discussed in \citet{saviane+2012}, we calculated  the ratio  $\rho$ between the r.m.s. dispersion around the linear fit with fixed slope {(here $\beta = -0.55$~\AA~mag$^{-1}$)}  
and the average measurement error in $\Sigma_{\rm EW}$.  
{This indicator of possible abundance spreads} is given in the last column in Tables~\ref{tab:cluster_paper2} and \ref{tab:cluster_paper1}. Clusters with $\rho \geq 2$ are considered candidates for a metallicity dispersion, while clusters with $1 < \rho < 2$ are marginal candidates.

\subsection{BH 176}
The metal rich cluster BH\,176 (\object{ESO 224-SC8}) is regarded as a transition stellar system, classified as either an old metal-rich open cluster or a young  metal-rich globular cluster with unclear origin \citep{ortolani+1995a,phelps+2003}. From isochrone fitting, \citet{frinchaboy+2006} found that the CMD is consistent with solar metallicity ($\rm[Fe/H]=0.0$ to $\rm[Fe/H]=+0.2$), with an age range from $5.6$ to $7.1$ Gyr. In a more recent analysis \citet{davoust+2011} found a metallicity  $\rm[Fe/H]=-0.1\pm 0.1$ and age $\sim 7$ Gyr by combining near-infrared 2MASS and optical photometry, a result confirmed by medium-resolution stellar spectroscopy by \citet{sharina+2014}. 

As shown in Fig.~\ref{clusters_all} and Table~\ref{tab:cluster_paper2}, after applying our selection criteria 11 stars were selected as members, with a relatively large dispersion in $\langle W^{\prime} \rangle$  ($0.43$~\AA) and a narrow radial velocity distribution ($\sigma_{v_{\rm r}} \approx 5$ km s$^{-1}$)
with average zero. This value agrees with previous measurements, in particular that of  \citet{sharina+2014}, and does not overlap the radial velocity of field stars expected from the Besan\c{c}on model \citep{robin+2003}, which is $v_{\rm r} =-71$ km s$^{-1}$ with a dispersion of $50$ km s$^{-1}$. 
We also tested the contamination of the sample by foreground dwarf stars  using the $\lambda 8806.8$ \AA\ Mg line. As shown by \citet{battaglia+2012},  the EW ratio between the Mg line and the two strongest CaT lines can be used to discriminate between dwarf and giant stars, due to the Mg line dependence on gravity. This method is particularly useful when the cluster and foreground samples have different metallicities \citep[see Fig.~7 in][]{dacosta+2014}. However, BH\,176 looks as metal rich as the Milky Way disk, which makes this comparison rather difficult.  From this method we identified two foreground dwarf stars. Excluding them does not change the mean metallicity of the cluster.

From our set of member stars we measured $\langle W^{\prime} \rangle =6.44$~\AA, which is the highest reduced equivalent width found in our sample, and   suggests a high metallicity in agreement with the photometric results.  
Applying the calibration of \citet{saviane+2012} or Eq.~\ref{V17S12scale} would require a large extrapolation (these relations are defined in the range $\sim1.8 < W^{\prime} < 5.8$)  and give an unrealistic iron abundance  $\rm[Fe/H] \geqslant +0.4$.
The less steep calibration of  \citet{vasquez+2015} yields a value for the BH\,176  mean metallicity  $\rm [Fe/H]=-0.12\pm0.22$, in excellent agreement with the results of full-spectrum fitting of  optical spectra by \citet{sharina+2014} and  \citet{dias+2016} ($\rm [Fe/H]=-0.08\pm0.04$). 
{Its location in Fig.~\ref{fig:Metallicity-values-from}, where it appears to be stronger lined than GCs of similar metallicity, suggests that  BH\,176 is an old metal-rich open cluster rather than a young metal-rich GC. 
}

%---------------------------------
\subsection{Djorgovski 1}

{This bulge cluster is one of the two faintest objects in our sample (together with Terzan 1), with a mean HB magnitude $V_{\rm HB}=21.15$. It is located at $(l,b)=(-3.33,-2.48)$, in a high-extinction area in the direction of the Galactic centre. 
NTT $V,I$ CMDs of  \object{Djorg\,1} by \citet{ortolani+1995b} gave a distance of 8.8 kpc and high reddening $E(B-V)=1.71$. The CMD was very dispersed, despite the good observing seeing.  \citet{valenti+2010}, by cleaning the CMD for contamination by the MW foreground, found a lower reddening and larger distance ([$E(B-V)=1.58$], $13.7$ kpc), and a metal-poor composition ($[\rm Fe/H]=-1.51$). 

In our study, the target selection from the CMD was rather difficult, and indeed only three stars have been classified as cluster members  after applying our selection criteria mainly driven by radial velocities. Our selection is consistent with the spatial distribution of the stars since all the rejected objects are located far from the cluster centre. From the member stars we obtained a mean metallicity  $\rm [Fe/H]_{S12}=-1.48$, consistent with the value derived from near-IR photometry \citep{valenti+2010}.
}
{We also confirm the exceptionally high radial velocity for a globular cluster ($-358$ km s$^{-1}$). Given the low galactic longitude and latitude of this cluster, the velocity is also high in the Galactic rest frame. 
Depending on the tangential velocity (proper motion), this might approach a space velocity close to the escape velocity from the Galaxy. }

\subsection{Djorgovski 2}

{The position of \object{Djorg 2} (also known as ESO456-SC38) close to the Galactic centre, $(l,b)=(2.7,-2.5)$ and $R_{\rm GC}=1.8$ kpc, makes extracting a clean CMD difficult because of crowding and high reddening [$E(B-V)=0.94$]. From NTT CMDs in the $V,I$ bands, \citet{ortolani+1997}  found a metallicity [Fe/H]~$ \sim -0.5$, while the near-infrared study of \citet{valenti+2010} gave  $\rm[Fe/H]=-0.65$, in both cases from the RGB morphology. \citet{rossi+2015} made an attempt to decontaminate the CMD using two-epoch observations and proper motions, which however did not give conclusive results because of a high differential reddening and the loose nature of the cluster.}

{As was the case with Djorg\,1, our selection criteria gave only three member stars with a mean radial velocity $v_{\rm r}\approx-160$ km s$^{-1}$, consistent with the velocity measured from our optical spectra \citep{dias+2016}, $v_{\rm r} =-150 \pm 28$ km s$^{-1}$.  The derived mean metallicity is $\mbox{[Fe/H]}_{\rm S12} = -1.09$, about 0.5 dex more metal-poor than the photometric determinations.  The metallicity from the optical region is $\mbox{[Fe/H]}_{\rm D16} = -0.79$, which is $\sim 0.3$ dex more metal rich than our present CaT estimate. 
}

{The optical CMD of Djorg\,2 obtained from our pre-imaging observations is shown in Fig.~\ref{djorg2_comp} together with two clusters with similar metallicity, NGC\,6401 and NGC\,6864. For each cluster we plot the photometry obtained from the inner cluster region ($r<0.5$ arcmin), on top of the full field photometry.  Over-plotted in each panel is the fiducial line published by \citet{catelan+2002} for NGC\,6864, which is helpful  to compare  the RGB and HB morphology. Given the little number of confirmed cluster members, this CMD does not discriminate between a higher or lower metallicity. A larger star sample and possibly high-resolution spectroscopy will be useful to pinpoint the metallicity of this cluster.
}

\subsection{\object{NGC\,6256}}

This globular cluster is located in the outer bulge at $(l,b)=(-12.21, 3.31)$ with a distance from the Galactic centre $R_{\rm GC}=3.0$ kpc. The high extinction of this cluster, $E(B-V)=1.01$,  prevented so far spectroscopic studies 
{of individual stars in the visible domain.} 
\citet{ortolonai+1999} analysed the optical $VI$  CMD, finding evidence of an extended blue horizontal branch (BHB) similar to other intermediate metallicity clusters with metallicity close to $\rm [Fe/H]=-1.5$. Integrated spectra from \citet{bica+1998} give a metallicity  $\rm [Fe/H]=-1.01$.
HST data analysis by \citet{piotto+2002} gave a $BV$ CMD (with a large photometric scatter) confirming the extended blue HB.   The adopted metallicity was relatively high, $\rm [Fe/H]=-0.70$. A lower metallicity is supported by the studies of \citet{stephens+2004} and \citet{valenti+2007}, who found $\rm [Fe/H]=-1.35$ and $\rm [Fe/H]=-1.63$,  respectively. 
The \citet{stephens+2004} measurements were performed on Na, Ca, and CO features using medium-resolution spectroscopy in the 2.2$\mu$m  spectral region, while the \citet{valenti+2007} value was obtained from near-IR photometry. 

We classified ten stars as cluster members, mostly because of their radial velocity, with only one star rejected because of its line strength. The mean radial velocity is consistent with previous estimates  ($\Delta v_{\rm r}=-2.0$ km s$^{-1}$ relative to H10). 
From those ten stars, the CaT line strengths favour a low metallicity, $\rm [Fe/H]_{\rm S12}=-1.61$. 
A high $\rho$ index  ($\rho=2.2$, the highest of the sample) suggests an intrinsic metallicity dispersion $\sigma_{\rm [Fe/H]} \approx 0.2$ dex, for which we plan a follow-up study.

\subsection{\object{NGC\,6355}}

This is another bulge globular cluster located $1.4$ kpc from the Galactic centre, in a relatively high-extinction area  [$E(B-V)=0.77$]  behind a dark nebula in the eastern part of the Ophiuchus complex. H10 and C09 used the \citet{zinn+1984}  analysis of integrated light as a reference metallicity, obtaining a mean metallicity  
$\rm [Fe/H] \approx -1.35$. 
More recent estimates  are from photometric studies in the optical \citep{ortolani+2003} and near-infrared \citep{valenti+2007} which suggest a metal poor value $\rm [Fe/H]=-1.6$, once  transformed to the C09 metallicity scale. Our metallicity measurement  from CaT is $\rm [Fe/H]_{\rm S12}=-1.59$, while \citet{dias+2016} found $\rm [Fe/H]_{\rm D16}=-1.46$.

\subsection{\object{NGC\,6366}}

This cluster has been catalogued as either a bulge  or a halo globular cluster \citep{campos+2013}. It is located at $3.5$ kpc from the Sun ($5$ kpc from the Galactic centre) and 1 kpc above the plane. Its radial velocity ($v_{\rm r}=-118$ km s$^{-1}$) is consistent with halo kinematics as pointed out by \citet{dacosta+1989}, who found $v_{\rm r}=-123$ km s$^{-1}$. The previous metallicity estimate listed in appendix 1 of C09, based on the CaT method applied by \citet{dacosta+1995} and \citet{dacosta+1989}, agrees with the value  measured by \citet{dias+2016} from optical spectra, $\rm [Fe/H]_{\rm D16}=-0.61$. 
{This value has recently been confirmed by detailed abundance analysis of eight red giant stars \citep{puls+2018} which gave [Fe/H]$= -0.60$ and a mean radial velocity $v_{\rm r}=-121.2$ km s$^{-1}$.}
Our present CaT estimate is $\rm [Fe/H]_{\rm S12}=-0.81 \pm 0.14$ (from 14 stars), that is  $\sim0.2$ dex less metal rich yet consistent within the error.  

{The photometric study of \citet{monelli+2013} suggested a bimodality on the red giant branch. We measured a relatively large $\sigma_{\rm [Fe/H]}$ ($\rho=1.3$) which marginally suggests an intrinsic metallicity dispersion. However, \citet{puls+2018} did not found any significant star-to-star abundance scatter.}
The projected radial distributions of stellar metallicities and radial velocities do not show any systematics with distance from the cluster centre.

\subsection{\object{NGC\,6401}}

This globular cluster is located in the inner bulge region at $\sim0.8$ kpc from the Galactic centre, with extinction $E(B-V)=0.72$. From optical photometry \citet{barbuy+1999} characterised this cluster as metal-rich, with metallicity similar to that of NGC\,104 ($\rm [Fe/H]=-0.70$). Later, the optical and NIR photometric studies of \citet{piotto+2002} and \citet{valenti+2007} detected a blue extended HB consistent with a more metal poor value  $\rm [Fe/H]=-1.56$. The presence of a blue HB is also confirmed by our optical photometry from the pre-imaging, and consistent with the metallicity derived here from seven cluster members ($\rm [Fe/H]_{\rm S12}=-1.30$).  
The metallicity obtained from optical spectra is $\rm [Fe/H]_{\rm D16}=-1.12$, in agreement with the only previous spectroscopic study of individual stars  ($\rm [Fe/H]=-1.1$, \citealt{minniti+1995}). 
%\citep[$\rm [Fe/H]=-1.1$][]{minniti+1995}. 
 
\citet{minniti+1995} obtained a radial velocity $-78\pm33$ km s$^{-1}$  while our value is $v_{\rm r}=-115$ km s$^{-1}$, 
{in agreement with the result of \citet{dias+2016}, $v_{\rm r}=-120 \pm 17$ km s$^{-1}$. This discrepancy  
is possibly related to field contamination in the \citet{minniti+1995} sample.} From the Besan\c{c}on model \citep{robin+2003} the typical radial velocity for this field is $v_{\rm r}\approx+60$ km s$^{-1}$ with a broad distribution ($\sigma_{v_{\rm r}}\approx 110$ km s$^{-1}$). These values are consistent with the velocity distribution of stars rejected from our sample, located at $r>2$ arcmin from the cluster centre.

\subsection{\object{NGC\,6426}}

This metal-poor halo globular cluster is located at $21$ kpc from the Sun in a relatively low-extinction area [$E(B-V)=0.39$]. The H10 catalogue lists a metallicity $\rm [Fe/H] = -2.15$ based on integrated spectra (ZW84). 
From CMD photometry, \citet{hatzidimitriou+1999} measured a metallicity $\rm [Fe/H]=-2.33 \pm 0.15$ 
{and an age marginally older than M\,92}. 
Such a low metallicity has been confirmed by \citet{dias+2015} ($\rm [Fe/H]_{\rm D16}=-2.39$) and more recently by high-resolution spectroscopy of four red giants by  \citet{hanke+2017}. Their detailed abundance analysis gives a mean Fe abundance $\rm [Fe/H] = -2.34 \pm 0.05$ dex with a mean $\alpha$-elements-to-Fe ratio of $0.39 \pm 0.03$. 
We classified here seven stars as cluster members from a relatively easy selection process. The target stars are clearly split in two groups: one with large negative radial velocities  and the other close to zero. This second group has radial velocities consistent with the Besan\c{c}on model predictions \citep{robin+2003}, as  shown in \citet{sharina+2012}.  For the members we found a mean radial velocity  $v_{\rm r}=-220$ km s$^{-1}$ which is more negative than the value published in H10 ($v_{\rm r}=-162$ km s$^{-1}$) and close to the result of \citet{hanke+2017} from high-resolution spectroscopy ($-212.2 \pm 0.5$ km s$^{-1}$). 
We measured a (slightly extrapolated) mean metallicity  $\rm [Fe/H]_{\rm S12}=-2.38$ for the cluster stars,  which is the lowest value in our sample and is consistent with the results of \citet{dias+2015} and  \citet{hanke+2017}.

\subsection{\object{NGC\,6517}}

{For this cluster, \citet{kavelaars+1995} found a metallicity [Fe/H]~$=-1.58\pm0.05$ from $BVI$ photometry after correction for differential reddening.  Previous optical low-resolution spectroscopy of individual stars by \citet{minniti+1995} gave  a metallicity $\rm [Fe/H]=-1.20$, close to the value derived from integrated light by ZW84, $\rm [Fe/H] = -1.34 \pm 0.15$.} From nine member stars we found  $\rm [Fe/H]_{\rm S12}=-1.63$,  a value significantly more metal-poor than 
{the metallicity tabulated by C09, which has its origin in the \citet{minniti+1995} and ZW84 measurements.}

\subsection{\object{NGC\,6864}}

NGC\,6864 (M\,75) is an outstanding globular cluster located $15$ kpc away from the Galactic centre in the interface region between the inner and outer Galactic halo \citep{zinn+1993,carollo+2007}. As one of the most massive and centrally concentrated GCs in the Milky Way \citep{harris+1996}, it contrasts with the typical characteristics of size and low concentration of GCs in the outer halo  \citep{koch+2009,koch+2010}.
Its relatively high metallicity derived from high-resolution spectroscopy  ($\mbox{[Fe/H]}=-1.16$, \citealt{kacharov+2013}),  together with an age of $\sim10$ Gyr \citep{catelan+2002}, have suggested a possible extragalactic origin.   
The spectroscopic study of \citet{kacharov+2013} found  three generations of stars apparently formed on a short timescale, more closely associated with an extended Na-O anticorrelation  than an Fe abundance scatter. The right panel of Fig.~\ref{djorg2_comp} shows an optical CMD derived from our pre-imaging photometry, confirming the HB morphology discussed by \citet{catelan+2002}, characterised by a tri-modal HB. 

From our measurements we estimate a cluster metallicity $\mbox{[Fe/H]}_{\rm S12} = -1.14$, consistent  with the estimate of \citet{dias+2016} ($\mbox{[Fe/H]}_{\rm D16} = -1.00$) and with previous measurements. 
A $\rho$ index of $1.3$  suggests a moderate metallicity spread from 11 member stars. This metallicity scatter, if not caused by a mix of stellar populations as suggested by  \citet{kacharov+2013}, can be related to the high luminosity, meaning mass, of this cluster ($M_{V}=-8.57$ mag), in accord with the correlation between metallicity dispersion and cluster luminosity found by \citet{carretta+2009}.

To directly compare our measurements with those of \citet{kacharov+2013} we looked for common targets, finding six stars. One of the six stars (labelled 1\_9285 in our programme) was rejected as a non-member by our selection   criteria due to its radial velocity $-168$ km s$^{-1}$. This value is 4$\sigma$ off with respect to the mean value measured from the 11 member stars ($v_{\rm r}=-193.4$ km s$^{-1}$, $\sigma_{v_{\rm r}}=5$ km s$^{-1}$). From the remaining five stars in common we obtain {mean values in good agreement:}  
$\rm [Fe/H]_{S12}=-1.13$ versus $\rm [Fe/H]_{HR}=-1.19$  (dispersion of the difference $\sim0.01$ dex).

\subsection{Pal 10}

\object{Pal 10} is a highly obscured globular cluster, with a reddening $E(B-V)=1.66$, located $0.3$ kpc above the Galactic plane and $6$ kpc away from the Sun. Its optical CMD shows significant effects from differential reddening, which introduces an uncertainty of the order $\sim0.2$ mag on the HB location.  A previous metallicity estimate comes from the optical photometry of \citet{kaisler+1997} who measured $\rm [Fe/H]=-0.1$. Our analysis gives a somewhat less metal rich value from 9 members stars, $\rm [Fe/H]_{S12}=-0.27$, to be compared to 
$\rm [Fe/H]_{D16}=-0.24$ from spectral synthesis in the optical range. The calibration of \citet{vasquez+2015} yields $\rm [Fe/H]_{V15}=-0.57$, illustrating the sensitivity to the different calibrations in the metal rich regime.

\subsection{Terzan 1}

The globular cluster \object{Terzan 1} is located almost on the Galactic plane ($0.1$ kpc above it), in the inner region of the Galactic bulge at $1.3$ kpc from the Milky Way centre.   Its reddening  $E(B-V)=1.99$ is the highest in our sample, with important differential effects on the optical CMD. This cluster is considered  one of the second-parameter globular clusters in the Galactic bulge (along with NGC\,6388 and 6441), because of the presence of a red HB along with a relatively low metallicity.  Previous metallicity estimates  include measures from optical photometry ($\rm [Fe/H]\approx-1.2$, \citealt{ortolonai+1999}), NIR photometry ($\rm [Fe/H]=-1.11$, \citealt{valenti+2010}),  and low-resolution optical spectroscopy from \citet{idiart+2002}  ($\rm [Fe/H]\approx-1.3$).  High-resolution near-infrared spectra of 15 stars in the inner region of Terzan\,1 \citep{valenti+2015} yield $\rm [Fe/H]=-1.26\pm0.03$ dex. 
Our selection provides 9 member stars, located in the inner region of the cluster ($r<2$ {arcmin}), 
with a mean velocity $v_{\rm r}=63$ km s$^{-1}$ and dispersion $\sigma_{v_{\rm r}}=8$ km s$^{-1}$.  This result is consistent with the result of \citet{valenti+2015}, $v_{\rm r}=57$ km s$^{-1}$, while there is a difference $\Delta v_{\rm r}=-51$ km s$^{-1}$ from the mean radial velocity measured  by \citet{idiart+2002}. 
From the mean $W^\prime$ of member stars we derived a metallicity $\rm [Fe/H]_{S12}=-0.84$, slightly more metal-rich than literature values {and the \citet{dias+2016eso} determination.}

\subsection{\object{Ton 2}}

The globular cluster Ton 2 (Tonantzintla 2) is located in the central region of the Galactic bulge ($R=1.4$ kpc), $0.5$ kpc below the Galactic plane. Its metallicity and extinction derived from NTT optical photometry are  $\rm [Fe/H]\approx-0.6$ and  $E(B-V)=1.24$, respectively \citep{bica+1996}.  Its HB morphology suggests an important differential reddening similar to that of Pal\,10, leading to an uncertainty of $0.22$ mag in the HB magnitude. {No previous spectroscopic observations exist for this cluster.}    
The CaT reduced equivalent width implies a high metallicity $\rm [Fe/H]_{S12}=-0.26$, which is also close to that of Pal\,10.  
{This value is more metal-rich than the values tabulated by H10 and \citet{dias+2016eso}, which originate from the CMD-based [Fe/H] determination of \citet{bica+1996}.}

\subsection{\object{Terzan 8}}

{Terzan 8 is a globular cluster in the Milky Way halo associated with the Sgr dwarf galaxy \citep{dacosta+1995,montegriffo+1998,law+2010}. The extinction along the line of sight is low, $E(B-V)=0.12$.} 
This cluster has previous measurements from high-resolution spectroscopy by \citet{mottini+2008} and \citet{carretta+2014}, giving metallicities  $\rm[Fe/H]=-2.35\pm0.05$ and $\rm[Fe/H]=-2.27\pm0.08$, respectively. From our measurements, we obtained 
$\rm [Fe/H]_{S12}=-2.15$, close to the results of high-resolution spectroscopy. From the optical spectra  the mean metallicity is  $\rm [Fe/H]_{D16}=-2.06$, in agreement with our CaT results within 1$\sigma$.

%__________________________________________________________________
\section{Summary and discussion}

\begin{figure}[t]
\centering
{\includegraphics[width=1.0\columnwidth]{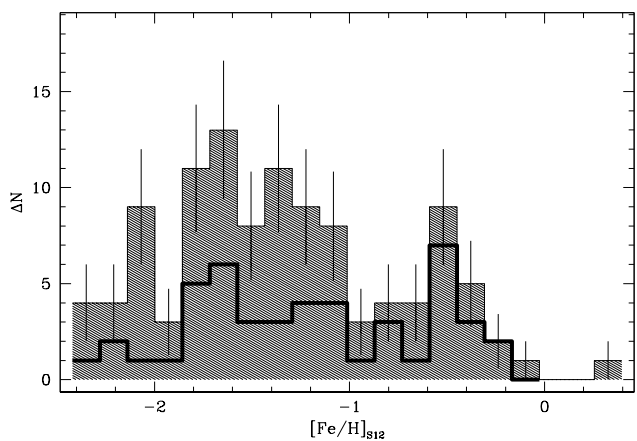}}
%{\includegraphics[width=1.0\columnwidth]{newplots/fhisto.h.eps}}
\caption{Metallicity distribution of the MW GCs in the merged CaT sample.  
The thick line shows the histogram made with our data only (Paper~I and the present work); the remaining data are from \citet{rutledge+1997} as put on the C09 scale by \citet{saviane+2012}.
The error bars are Poisson errors on the number of clusters in each bin.}
\label{fhisto}
\end{figure}

We have obtained new radial velocities and metallicities for stars in 28 Milky Way GCs, {as the completion of our spectroscopic survey} aimed at providing a  homogeneous set of metallicities from medium resolution data in two spectral regions. The results in this paper are based on FORS2 spectroscopy with a resolution $R\sim2500$ in the region centred on the \ion{Ca}{II} triplet lines ($8450-8700$ \AA) {of individual cluster candidate red giants.}
In a previous study of this series,  \citet{saviane+2012} measured the metallicity of 20 programme clusters using eight template clusters with known metallicity to establish an empirical CaT line strength calibration. 
In this paper we have measured the metallicity of an additional sample of 28 globular clusters mainly in the Galactic bulge or far in the halo.  Most of the clusters analysed here are therefore faint and/or highly reddened, which for a long time prevented any metallicity measurement based on spectroscopy of individual stars. 
{We provide metallicities based} on the CaT calibration of \citet{saviane+2012}, with typical uncertainties of the order $\sim0.15$ dex. We also discussed the impact on the measured [Fe/H] of alternative empirical calibrations such as those of \citet{vasquez+2015} and \citet{dacosta+2016}, with differences especially noticeable in the metal-rich regime (${\rm [Fe/H]>-0.8}$). 
{Our sample is complemented by the 71 GCs published by \citet{rutledge+1997} transformed to our metallicity scale} \citep{saviane+2012}.  
Altogether,  our project provides the largest GC sample with homogeneous metallicity measurements from CaT currently available. 
The combination of all data yields CaT metallicity estimates  on the same scale for
{107 GCs, that is  ${{\sim69\%}}$} of the clusters published in H10.
The metallicity distribution of all clusters in our CaT project is plotted in Fig.~\ref{fhisto}, where the new data represent about 50\% of the available measurements  for metal-rich clusters.
As a test of the metallicity scale provided by the \ion{Ca}{ii} triplet, we compared our results with those obtained from spectroscopy of stars in the same clusters using a spectral fitting technique in the optical region  \citep{dias+2015,dias+2016}. The comparison shows that the metallicities measured  from the CaT line strengths using the calibration of \citet{saviane+2012} 
are consistent within the errors with the [Fe/H] values derived from spectral fitting by Dias et al.,  which in turn agree with metal abundances {from high-resolution spectroscopy.} 
Our study confirms the CaT method as a powerful tool to determine [Fe/H]  for stellar systems so distant or reddened that high-resolution spectroscopic analysis becomes very difficult or prohibitive, provided that the calibrators are selected from the same population as the targets. {This will be particularly useful to study resolved stars beyond the Local Group with the new generation of extremely large telescopes.} 

%--INDIVIDUAL CASES:
We have identified some interesting cases among the 28 GCs analysed in the present work.  We obtained a much improved metallicity measurement for Djorg\, 2: while previous work based on a NIR CMD classified it as a metal-rich globular cluster with a metallicity ${\rm [Fe/H]=-0.65}$,  our EW measurements and optical photometry suggest a more metal-poor value ${\rm [Fe/H]\approx-1.1}$. Additionally, we found a radial velocity $v_{\rm r}=-160$ km s$^{-1}$ which is the first estimate for this cluster. Another interesting case is that of Terzan 1, for which we found a difference $\Delta v_{\rm r}=-51$ km s$^{-1}$ in  radial velocity with respect to previous values.  Our estimate is consistent with the latest measurement from high-resolution spectra \citep{valenti+2015}. Finally, as in \citet{saviane+2012}, we explored the presence of intrinsic metallicity dispersions within our sample: we found three clusters that might have a metallicity dispersion, based on a comparison of their EW dispersion with the intrinsic error in EW measurements (parameter $\rho\geqslant1.3$). One of them, NGC\,6256, is the most probable candidate, with $\rho=2.2$. For this cluster our data suggest an intrinsic metallicity dispersion $\sigma_{\rm [Fe/H]}=0.2$ dex, but more data are needed to confirm this hint.

%______________________________________________________________

\begin{acknowledgements}
{The authors thank the anonymous referee for helpful comments and suggestions.
SV and MZ acknowledge support   by  the  Ministry  of  Economy,
Development, and Tourism's Millennium Science Initiative through grant
IC120009, awarded  to The Millennium Institute  of Astrophysics (MAS),
by  Fondecyt  Regular  1150345
and   by  the  BASAL-CATA  Center  for
Astrophysics and Associated Technologies PFB-06. 
SV also acknowledges support by the ESO studentship programme 2012-2014. }
\end{acknowledgements}

\bibliographystyle{aa} % style aa.bst
\bibliography{svasquez}

%--------------------------------------------------------------------------------------------------------------------------
\end{document}